\documentclass[11pt]{article}

\usepackage[margin=1in]{geometry}

\usepackage{longtable}
\usepackage{booktabs}
\usepackage{multirow}
\usepackage{array}
\usepackage{microtype}
\usepackage[round]{natbib}
\usepackage{graphicx}
\usepackage{float}
\usepackage{placeins}
\usepackage{pdflscape}
\usepackage{amssymb} 
\usepackage{enumitem}

\usepackage{caption}
\usepackage{titletoc}
\usepackage{makecell}

\usepackage{tikz}
\usetikzlibrary{positioning, arrows.meta}
\usepackage[most]{tcolorbox}

\usepackage[colorlinks=true, linkcolor=blue, citecolor=blue, urlcolor=blue, breaklinks=true, pdfborder={0 0 0}]{hyperref}
\newcommand{\okemoji}{\includegraphics[height=1em]{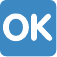}}
\newcommand{\warnemoji}{\includegraphics[height=1em]{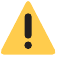}}
\newcommand{\xemoji}{\includegraphics[height=1em]{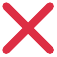}}

\begin{document}
\title{Measurement to Meaning: A Validity-Centered Framework for AI Evaluation}

\author{
    Olawale Salaudeen\textsuperscript{1}\thanks{Equal contribution.}~\thanks{Corresponding authors: \texttt{olawale@mit.edu; sanmi@cs.stanford.edu.}} \and
    Anka Reuel\textsuperscript{2}\footnotemark[1] \and
    Ahmed Ahmed\textsuperscript{2} \and
    Suhana Bedi\textsuperscript{2} \and
    Zachary Robertson\textsuperscript{2} \and
    Sudharsan Sundar\textsuperscript{2} \and
    Ben Domingue\textsuperscript{2} \and
    Angelina Wang\textsuperscript{2,3}\thanks{Equal senior authorship.}\and
    Sanmi Koyejo\textsuperscript{2}\footnotemark[3]~\footnotemark[2]
}

\date{}
\maketitle

\begin{center}
\begin{tabular}{c c c}
\textsuperscript{1} Massachusetts Institute of Technology &
\textsuperscript{2} Stanford University &
\textsuperscript{3} Cornell Tech
\end{tabular}
\end{center}

\begin{abstract}
    While the capabilities and utility of AI systems have advanced, rigorous norms for evaluating these systems have lagged. Grand claims, such as models achieving general reasoning capabilities, are supported with model performance on narrow benchmarks, like performance on graduate-level exam questions, which provide a limited and potentially misleading assessment. We provide a structured approach for reasoning about the types of evaluative claims that can be made given the available evidence. For instance, our framework helps determine whether performance on a mathematical benchmark is an indication of the ability to solve problems on math tests or instead indicates a broader ability to reason. Our framework is well-suited for the contemporary paradigm in machine learning, where various stakeholders provide measurements and evaluations that downstream users use to validate their claims and decisions. At the same time, our framework also informs the construction of evaluations designed to speak to the validity of the relevant claims. By leveraging psychometrics’ breakdown of validity, evaluations can prioritize the most critical facets for a given claim, improving empirical utility and decision-making efficacy. We illustrate our framework through detailed case studies of vision and language model evaluations, highlighting how explicitly considering validity strengthens the connection between evaluation evidence and the claims being made.
\end{abstract}

\section{Introduction}
\begin{figure}[t]
    \centering
    \includegraphics[width=\textwidth]{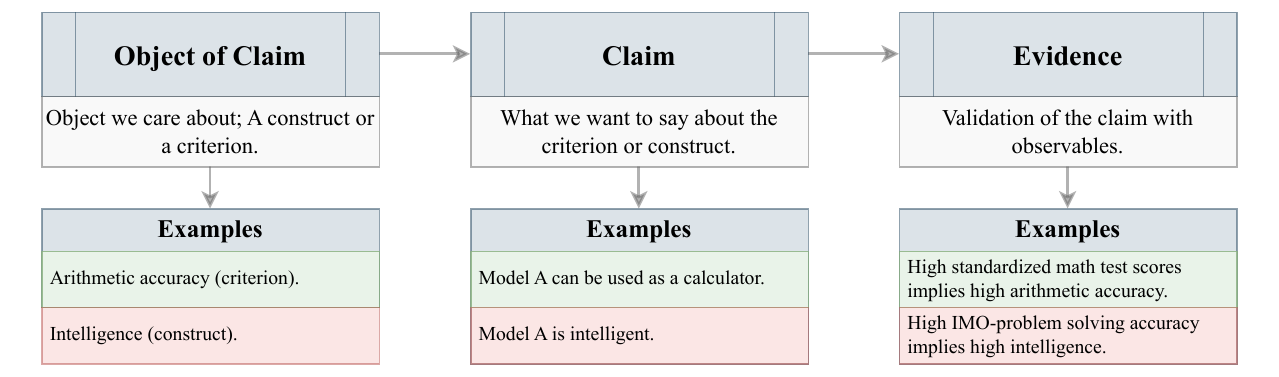}
    \caption{The three components are needed to begin the process of validation. First, we must decide what object our claim is about; is it a criterion, or is it a construct? Then, we must explicitly state the claim. Finally, we must identify or design our evidence and assess whether it supports the desired claim, i.e., do we have a valid claim based on the evidence? Here, a green background indicates that the claim-evidence pair is reasonably well supported. In contrast, a red background means the inferential leap between claim and evidence is larger and less well-supported.}
    \label{fig:three_components}
\end{figure}

\begin{table}[t]
    \centering
    \caption{Table of terms and definitions.}
    \label{tab:term_defs}
    \resizebox{\textwidth}{!}{
    \begin{tabular}{|l|p{5cm}|p{5cm}|p{5cm}|}
\hline
\textbf{Term} & \textbf{Definition} & \textbf{How does it relate to other terms?} & \textbf{Example} \\
\hline
Measurement Instrument & A tool used to gather observations or assign values (e.g., a benchmark, user study, or survey). & Underlies the act of measurement. Evaluation and claims often hinge on data obtained via instruments. & A dataset of IMO math problems (the “instrument”) used to gather system accuracy scores. \\
\hline
Measurement & Assigning a quantitative or qualitative value to a property of a system (e.g., accuracy, usability). & Involves applying the instrument and recording results; informs subsequent evaluation. & “The system answers 15 of 20 IMO questions correctly” (accuracy = 75\%). \\
\hline
Evaluation & The broader process of interpreting one or more measurements in context. & Translates raw measurement into insights (e.g., domain-specific analysis, comparisons to a baseline). & “Because the system can solve 75\% of these IMO problems, it demonstrates proficiency in competition-level algebra.” \\
\hline
Claim & An assertion, judgment, or decision made about the system, potentially based on evaluation results. & Draws on evaluation evidence to generalize or conclude something about the system or its capabilities. & “The system exhibits human-level math reasoning skills.” \\
\hline
Criterion & A directly measurable or observable concept (e.g., ``textbook linear algebra question-answering accuracy''). & Can be measured directly and often serves as a baseline or gold standard for evaluation. & ``Textbook Linear Algebra question-answering accuracy'' – the system’s performance on textbook linear algebra questions. \\
\hline
Construct & An abstract concept not directly measurable (e.g., ``mathematical reasoning'' or ``trustworthiness''). & Requires an operational definition plus proxies or indicators to measure and evaluate indirectly. & ``Mathematical reasoning'' – a theoretical ability captured through various problem sets and expert assessments. \\
\hline
\end{tabular}
    }
\end{table}

\begin{table}[t]
    \centering
    \caption{We provide an overview of the different forms of validity considered in this work, along with key questions to ask in their assessment. The standard of evidence for validity depends on the conceptual gap between the measurement and the object of the claim, with broader gaps demanding stronger justification. Certain forms of validity, such as criterion validity, encompass multiple facets that capture different aspects of the evaluation. We adopt a view on validity closest to~\citep{lissitz2007suggested}. This includes aspects~\cite{Cronbach1955-sc} and \cite{Messick1995-vu}'s views on validity. Like~\citep{lissitz2007suggested}, we do not unify all facets of validity under construct validity.}
    \label{tab:validity_defs}
    \resizebox{\textwidth}{!}{
    \begin{tabular}{|l|l|p{7cm}|p{7cm}|}
\hline
\multicolumn{2}{|l|}{\bf Validity Type} & \textbf{Description} & \textbf{Example: IMO Problem Solving $\rightarrow$ Reasoning} \\
\hline
\multicolumn{2}{|l|}{\bf Content Validity} & Does your evaluation cover all relevant cases? & Does solving IMO problems sufficiently capture the content relevant to reasoning? \\
\hline
\multicolumn{2}{|l|}{\bf Criterion Validity} & Does your evaluation correlate with a known validated standard? & Does IMO problems accuracy predict other external criteria of reasoning, e.g., common sense reasoning benchmarks? \\
\hline
& Predictive Validity & Can your evaluation predict downstream outcomes? & -- \\
\hline
& Concurrent Validity & To what extent does your evaluation agree with another validated assessment under the exact same conditions? & -- \\
\hline
\multicolumn{2}{|l|}{\bf Construct Validity} & Does your evaluation truly measure the intended construct? & Does IMO problem solving capture all components of reasoning and only components of reasoning? \\
\hline
& Structural Validity & Does your evaluation capture the structure of the construct you are measuring? & -- \\
\hline
& Convergent Validity & Does your evaluation correlate with other measures that assess the same construct? & -- \\
\hline
& Discriminant Validity & Can your evaluation differentiate between constructs that should be distinct? & -- \\
\hline
\multicolumn{2}{|l|}{\bf External Validity} & Does your evaluation generalize across different environments or settings? & Does excelling at IMO problems translate to solving textbook linear algebra problems accurately, where the problems are provided in different formats? \\
\hline
\multicolumn{2}{|l|}{\bf Consequential Validity} & Does your evaluation consider the real-world impact of test interpretation and use? & Does emphasizing IMO problem-solving in AI development narrow research focus in ways that overlook other essential reasoning skills? \\
\hline
\end{tabular}
    }
\end{table}

Suppose we observe that an AI system can solve International Math Olympiad (IMO) problems accurately~\citep{Glazer2024-fy}. Let's consider two claims about what this ability implies:

\begin{enumerate}[label=\textbf{Claim \arabic*.}, align=left, labelwidth=0pt, labelsep=1em]
    \item The system can also solve linear algebra questions from a textbook accurately.
    \item The system has reached human-expert-level mathematical reasoning.
\end{enumerate}

Clearly, asserting Claim 2 requires a much greater inferential leap from the observed evidence (IMO accuracy) than Claim 1. If we claim that good performance on IMO problems demonstrates competence in solving linear algebra questions from a textbook, the justification may be reasonable: IMO problems often involve advanced undergraduate-level techniques, including linear algebra, so proficiency in them provides reasonable evidence for the claim. However, if we claim that the system has reached human-level reasoning, the justification is much weaker. Solving IMO problems primarily requires mathematical problem-solving, but human reasoning encompasses a broader spectrum, including common sense, adaptability, and metacognition, which IMO performance says little about. This difference highlights that we must scrutinize an evaluation and measurement in the context of the claim we wish to support. Thus, we consider validity. Validity refers to the degree to which evidence and theory support the interpretations of test scores for proposed uses of tests~\citep{aera2014standards}.

Historically, measurement theory studies how latent psychological concepts are quantitatively assessed, ensuring that the chosen measures accurately capture the intended constructs while maintaining validity and reliability. {\em Validity is not inherently a property of the measurement itself—it also depends on the context of the evaluation it enables, the claims supported by that evaluation, and the potential real-world consequences of those claims}~\citep{Messick1998-nb,Messick1995-vu, Shepard1993-cc, lissitz2007suggested, Cronbach1955-sc}. While this view of validity is not universal (c.f., ~\cite{Borsboom2004-zt}), it is widely held and endorsed by several professional organizations focused on measurement in educational and psychological testing~\cite{aera2014standards}. We discuss validity and its treatment in other scientific disciplines in Appendix~\ref{sec:validity}.

We first clarify some terms (we refer the reader to also carefully examine Table~\ref{tab:term_defs}). In this work, measurement refers to assigning a quantitative or qualitative value to a specific property of a system (e.g., accuracy or usability), while evaluations are the broader process of interpreting these measurements to provide insights about the system. Measurements and evaluations can then support claims, judgments, assertions, and decisions about the system. Measurement instruments are tools to gather observations or assign values and include benchmarks, user studies, and expert assessments.

For instance, we can measure accuracy on question-answer problems. However, the context of applying this measurement to IMO problem-solving problems (our measurement instrument) makes it an evaluation of a system’s accuracy in answering IMO problem-type questions; we are not merely recording an accuracy score but interpreting it in a specific domain context (IMO problems) to gain some insight about the system’s capabilities. To emphasize the point, while the measurement is an accuracy score, the interpretation of that measurement as an indicator of math problem-solving capability is an evaluation. Finally, one may then make claims---not necessarily correctly---about general reasoning capabilities to be supported by the evaluation.

For another example, we can measure the frequency of harmful outputs (e.g., misinformation or offensive responses) from a language model. The context of applying this measurement specifically to high-stakes medical advice scenarios (our measurement instrument) transforms it into an evaluation of the system’s safety in that domain; we are not merely counting harmful responses but interpreting their potential impact within a clinical setting. From there, one might make broad claims about the system’s trustworthiness or readiness for real-world deployment—claims that may be more or less justified depending on whether the measurement truly captures the range of potential harms and aligns with relevant medical standards. This full pipeline is relevant context for establishing validity.

We can measure without evaluating—for example, by collecting raw accuracy scores without concluding their implications. However, to evaluate, we must measure in some form (quantitatively or qualitatively) and then interpret those measurements in a domain-specific context. One might then ask, why measure if not to evaluate a system? We can measure as a means to develop new metrics for future evaluation; we can measure to observe or characterize phenomena before making judgments; we can measure for calibration. We may measure accuracy to identify patterns or outliers that guide future studies without evaluating whether the system’s performance is `good' or meets real-world requirements.

A core limitation of current AI evaluation discourse is that validity, if considered, often focuses on the measurement–evaluation relationship, i.e., designing measurements that support a predefined evaluated object~\citep{Gema2024-ja, Wallach2025-hx}, an important aspect of validity. Consequently, one may then conclude that if a measurement does not fully meet the needs of the object it was designed to evaluate, then no valid claims can be made, and no insights can be gained; but, just because IMO accuracy does not sufficiently measure reasoning does not mean that it cannot support better-scoped claims.

Additionally, {\em establishing validity is an iterative process}~\citep{Cronbach1955-sc,Kuhn1997-mx}, {\em there is no single validity checklist or iteration that can be completed to resolve all issues of validity.} Indeed, tests for human intelligence have existed for over a century~\citep{neisser1976cognition, binet1916development}, and the quest to improve the validity of tests is ongoing. Recognizing the limitations of measurements and evaluations, rather than outright rejecting them when certain validity criteria fall short, requires nuance; our framework enumerates this nuance. Our approach is essential for practical utility, enabling us to extract meaningful claims even from evaluations that do not rigorously satisfy all conditions of validity.

To assess the validity of a claim derived from an evaluation, Figure~\ref{fig:three_components}, we must:
\begin{enumerate}
    \item Carefully consider the object of the claim. Is it a construct—an abstract object that cannot be measured directly, like `mathematical reasoning'? Or is it a criterion—a directly measurable object, such as ``accuracy on linear algebra questions from a specific textbook''?
    \item Furthermore, does the claim refer to the same property measured, or does it extend beyond the specific evaluation to infer something about a different property? For example, one might measure IMO accuracy as part of an evaluation of mathematical ability, then use this evaluation to support a claim about textbook linear algebra question-answering, which is a different object.
    \item Finally, is the claim supported directly by the measurement (e.g., IMO accuracy implies textbook linear algebra question-answering capability), or does it rely on an intermediate construct (e.g., IMO accuracy implies mathematical reasoning, which in turn implies  textbook linear algebra question-answering capability)?
\end{enumerate}

These distinctions determine the necessary standards of evidence required before an evaluation can meaningfully support a claim. The alignment between what is measured, how it is interpreted (evaluation), and the overarching claim is central to establishing validity.

Ensuring validity primarily requires five forms of validity from psychometrics~\citep{Thorndike1949-nv, Cronbach1955-sc, Messick1995-vu, aera2014standards, lissitz2007suggested}, outlined in Table~\ref{tab:validity_defs}. Additionally, Section~\ref{sec:establishing} and Appendix~\ref{sec:validity} Table~\ref{tab:validity} enumerates tools to investigate and establish each form. While other forms of validity exist and are relevant\footnote{Other forms of validity include, for example, face validity. Additional forms of validity are given in~\citep{Lim2024-kw, Hughes2018-nd, Borsboom2004-zt, aera2014standards, hunsley2003incremental, reichardt2002experimental, bronfenbrenner1977toward, neisser1976cognition}.}, we identify this set as most relevant for current AI measurement validity gaps in Section~\ref{sec:gaps}. As part of our practically guided framework, we also differentiate between types of claims, where some forms of validity are trivially satisfied and only a subset needs to be focused on~\citep{lissitz2007suggested}.

The standard of evidence required to demonstrate validity depends on the conceptual gap between what is actually measured (and how it is evaluated) and the object of the desired claim. The greater the gap, the more arduous the task of establishing validity. Notably, different forms of validity are not independent and work together to demonstrate validity—we illustrate this in our case studies in Section~\ref{sec:framework} and Appendix~\ref{sec:casestudies}. Our proposed framework ties claims directly to the requisite standard of evidence provided by the evaluation. This alignment ensures the appropriate downstream use of evaluations while also guiding improvements that support more general claims.

This framework is pivotal because AI evaluations inform decisions with real-world consequences. For example, under Article 51 of the EU AI Act (5), benchmarks are explicitly referenced as indicators for classifying AI models according to their systemic risk. Developers of models deemed high-risk must comply with significantly more obligations. If we fail to consider validity in this context, such classification may become meaningless because we cannot be certain that what truly matters—namely, the risk posed by these models—is accurately captured by the chosen measurement instruments (e.g., benchmarks). This can lead to a false sense of security. Similarly, measurement instruments are often used within organizations~\citep{Hardy2024-ax} to guide resource allocation and further training aimed at improving a model’s capabilities. Yet, if the chosen instrument does not accurately measure the capability developers care about, additional training may simply become an exercise in ``teaching to the test''~\citep{Jennings2014-kc} rather than leading to genuine improvements in the model.

Recognizing these challenges, we propose a structured framework to assess the validity of AI assessments, ensuring appropriate use and interpretation. Specifically, {\bf our contributions} are: 

\begin{enumerate}
    \item We examine and identify limitations in how the relevance of different forms of validity has co-evolved with the progress of  AI and the corresponding norms and practices of evaluation. (Sec.~\ref{sec:gaps})
    \item We enumerate common risks to validity and practical operationalization to investigate and support distinct forms of validity. (Sec.~\ref{sec:establishing})
    \item We propose a practical and structured claim-aware framework for identifying the necessary evidence to establish the validity of claims based on AI evaluations. We also enumerate adoptable practices to demonstrate validity. (Sec.~\ref{sec:framework})
    \item We illustrate our framework through vision and language evaluation case studies, providing concrete, prescriptive examples of validating claims based on evaluations. (Sec.~\ref{sec:application} and App.~\ref{sec:casestudies})
\end{enumerate}

\section{Validity Gaps in Current AI Evaluations and Related Work} \label{sec:gaps}
AI evaluation has evolved alongside the systems it measures, but the distance between what we evaluate and what we claim about real-world utility has grown (Appendix~\ref{sec:evolution}), revealing gaps across all five major forms of validity, content, criterion, construct, external, and consequential (Table~\ref{tab:validity_defs}). These gaps reflect shifts not only in how we evaluate but in what we claim.

Initially, evaluations emphasized held-out test data drawn from the same distribution as training data, which is often called the i.i.d. setting, supporting content validity, where the test environment closely matched the training one. As pretraining became the norm, evaluation shifted to fine-tuning large models like those trained on ImageNet~\citep{Deng2009-xh, Russakovsky2014-ka, Kornblith2018-ho, Recht2019-at} and measuring downstream performance. This introduced criterion validity, where success on downstream tasks was taken as evidence that pretraining had captured useful representations~\citep{mikolov2013efficient, brown2020language, radford2021learning}.

As concerns about spurious correlations driven decision-making, distribution shifts~\citep{bai2025explicitly, Salaudeen2024-rt, Lopez-Paz2016-zr, Xiao2020-oe, Arjovsky2019-og, Rosenfeld2020-bb, Koh2020-qd, Gulrajani2020-bc}, and causal representations~\citep{scholkopf2021toward, gichoya2022ai} rose, so too did other forms of validity. Recent work has increasingly emphasized external validity, whether evaluation results generalize beyond the training distribution~\citep{Salaudeen2024-yu}, consequential validity, whether model deployment leads to desirable outcomes~\citep{Wang2023-nf}, and construct validity, whether an evaluation actually captures the concept we think it does~\citep{bell2024reassessing, salaudeen2025domain}.

Still, evaluation remains largely benchmark-driven, often in service of leaderboard-based progress \citep{Hardt2021-gy, Orr2024-sx}. This kind of evaluation is not without merit: when optimizers, architectures, or training procedures improve benchmark performance across multiple tasks, they also tend to improve performance on another new and real-world task, a sort of criterion validity~\citep{Blum2015-oc, Kornblith2018-ho, Salaudeen2024-yu}. Moreover, these shared benchmarks have helped align academia, industry, and other stakeholders on a criterion to measure progress~\citep{Donoho2023-tm, recht2024mechanics}, for instance, ImageNet accuracy~\citep{Russakovsky2014-ka}. However, benchmark performance does not always translate to reliable real-world performance or trustworthy decision making~\citep{Hardy2024-ax}.

The rise of foundation models, which can operate across diverse tasks out of the box, further complicates this issue. Traditional evaluation methods increasingly fail to capture real-world AI behaviors that require investigating abstract capabilities like intelligence and reasoning to predict broad and diverse downstream utility~\citep{Wu2023-ck,Wan2024-eu,Mirzadeh2024-as}.

Narrow datasets used for ``general-purpose'' evaluation raise content, construct, and external validity concerns, especially for constructs like reasoning~\citep{Bostrom2020-ps, alaa2025medical}. Furthermore, these evaluations lack criterion validity and fail to predict criteria of real-world utility~\citep{Hardy2024-ax}, and the socio-technical gap between evaluation results and real-world needs undermines consequential validity~\citep{Liao2023-nf, barocas2023fairness}. Consequently, overgeneralized results erode evaluation credibility~\citep{Raji2021-iv}.

Important prior work has demonstrated the need for validity frameworks~\citep{Jacobs2021-no,Saxon2024-fi,Subramonian2023-lw,Xiao2023-og,Blodgett2021-gj,Coston2023-rt,Xiao2024-aa,Reuel2024-tc}. However, much of it has focused on validity in the context of the limitations of measurements, focusing on conditions for perfect measurements of nicely defined concepts, which is far from practice. METRICEVAL~\citep{Xiao2023-og} raises validity concerns stemming from vague benchmark articulation and repurposed datasets in 16 natural language generation metrics. \cite{Liu2024-pi} further challenged benchmarks’ ability to measure intended constructs, proposing the Evidence-Centered Benchmark Design (ECBD) framework to ensure rigorous metric selection. However, these works focus on designing new measurement instruments for evaluations. While developing better measurement instruments for better evaluations is also important, and our framework also applies to this task, we find it of practical value to understand what claims can be made from existing evaluations and evidence, given the intractability of creating tailored evaluations for each claim, and the already unwieldy amount of existing benchmarks~\citep{Unknown2024-mm}.

The important work of~\citep{Chouldechova2024-im,Wallach2025-hx} applied~\citep{Adcock2001-lc}’s measurement theory, critiquing ML evaluations for conflating systematizing a background (conceptual and informal) concept with operationalizing (measuring) a systemized (formalized and structured) concept\footnote{According to Adcock and Collier, a background concept is a “broad constellation of meanings and understandings associated with [the] concept,” and systematization describes the process of refining and explicitly defining a concept to create a structured and consistent foundation for measurement and analysis (the systematized concept) while operationalization the process of transforming a systematized concept into measurable indicators~\citep{Adcock2001-lc}.}.

Our work complements this literature by explicitly identifying that validity depends not only on the measurement and evaluation but also on the claim intended to be made. Building on~\cite{Wallach2025-hx}—who underscore the importance of explicit systematization needed in AI, where concepts often emerge from practice rather than theory—our work clarifies this process in the context of nomological networks~\citep{Cronbach1955-sc}. These networks represent not only the relationship between the background concept and systematized concept but also the broader relationships to other background and systematized concepts. In the sense of the Duhem-Quine thesis\footnote{The Duhem-Quine thesis emphasizes that scientific claims are interconnected, meaning that rejecting or modifying one hypothesis affects others within the theoretical framework.}, a nomological network serves as a map of empirical and theoretical relationships, helping manage the holistic nature of scientific testing by clarifying how constructs relate to each other as well as observable evidence. Consequently, we expand upon~\cite{Wallach2025-hx}’s view of systematization by arguing for the importance of broader nomological networks beyond just locally specifying which definition of a background concept will be used. Our view subsumes both the view that background concepts (constructs) merely serve as a mechanism for conceptual agreement and the view that they are fundamental properties inherent to AI systems to be discovered and refined through the scientific process. This process must be considered in the emerging science of AI evaluations~\citep{weidinger2025toward, hardt2025emerging}.

Ultimately, our framework takes a practical approach, emphasizing that validity is not only a property of measurement and evaluation. As Cronbach and Meehl emphasize~\citep{Cronbach1955-sc}: ``In one sense, it is naive to inquire ‘Is this test valid?’ One does not validate a test, but only a principle for making inferences. If a test yields many different types of inferences, some can be valid and others invalid.'' We further enumerate risks, tools, and evidence exemplars to assess whether evaluations meet appropriate validity standards in Section~\ref{sec:establishing} and Table~\ref{tab:validity}.

\section{Risks to Validity and Operationalizable Strategies for Mitigation}\label{sec:establishing}
In this section, we examine common risks to valid claims as a function of limitations in establishing some form of validity and discuss existing tools and methodologies for assessing and strengthening validity in AI assessment. Appendix~\ref{sec:evidence} Table~\ref{tab:validity} categorizes in detail key risks, investigation tools, and evidence exemplars across multiple forms of validity in assessment. We summarize here. Importantly, this section makes clear that general-purpose {\em `benchmarks' are currently an insufficient sole evaluation mechanisms for the real-world utility of AI systems.}

Risks to content validity include coverage deficiency, where important aspects of the construct are missing, and construct irrelevance, where extraneous factors influence scores~\citep{aera2014standards, Messick1995-vu}. Imbalanced content can lead to assessments overemphasizing certain skills while neglecting others. These issues can be examined through expert review, adversarial scrutiny, and synthetic data generation, with supporting evidence from explicit content mapping and coverage analysis.

Risks to external validity include sample bias, where the test is validated on a narrow or unrepresentative population~\citep{Henrich2010-zc}, and unrealistic testing conditions, which may not reflect real-world scenarios~\citep{Donald-T-Campbell1963-up}. Temporal variability and interaction effects can also distort results if performance shifts over time or due to specific environmental factors~\citep{Andonov2023-fs}. These issues can be investigated through stress testing, A/B testing, transfer testing, and population-stratified assessments, with evidence from performance comparisons across different conditions and sensitivity analyses.

Risks to criterion validity include criterion contamination, where extraneous factors influence assessment, and criterion deficiency, where relevant aspects of performance are omitted~\citep{Brogden1950-ge,Austin1992-ta}. Restricted range limits the ability to detect meaningful relationships if all scores are too similar. These issues can be addressed through real-world longitudinal studies, validated criterion studies, and behavioral testing, with evidence from correlations with gold-standard benchmarks and predictions of real-world utility.

Risks to construct validity can come from structural, convergent, and discriminant validity risks. Structural validity is compromised by poor factor structure, where test items fail to group in expected ways~\citep{Clark1995-pq,Elhami-Athar2023-ql}, and complex measurement range, where constructs are not well captured across different levels of ability~\citep{Messick1995-vu}. Convergent validity can suffer from high measurement error~\citep{Cheung2024-gx}, which reduces reliability, while discriminant validity can be compromised by construct overlap, where different abilities are not clearly distinguished~\citep{Shaffer2016-zk}. These risks can be investigated using hypothesis testing, factor modeling, and benchmark suites, with supporting evidence from item-test correlations and demonstrated non-significant overlap with unrelated constructs.

Risks to consequential validity include bias and fairness issues, where results systematically disadvantage certain groups~\citep{meredith1993measurement, Messick1995-vu, Randall2023-jc}. While bias and fairness can themselves be constructs of interest, they are also important to consider in any measurement. Further, unintended incentives can distort behavior if assessment criteria encourage gaming rather than genuine learning~\citep{Nichols2007-jq}. Policy consequences may emerge if flawed assessments influence high-stakes decisions. These risks can be assessed through anticipatory ethics methods~\citep{Umbrello2023-dv}, societal impact audits, and ethical stress testing, with evidence from stakeholder feedback, improvements in fairness and reliability, and documented real-world impacts.

While this framework highlights key risks and mitigation strategies, additional risks may arise in different contexts, necessitating continuous assessment and refinement.

\section{A Framework for Claim-Centered Validity Assessment in AI Evaluation} \label{sec:framework}
\begin{figure}[t]
    \centering
    \includegraphics[width=\linewidth]{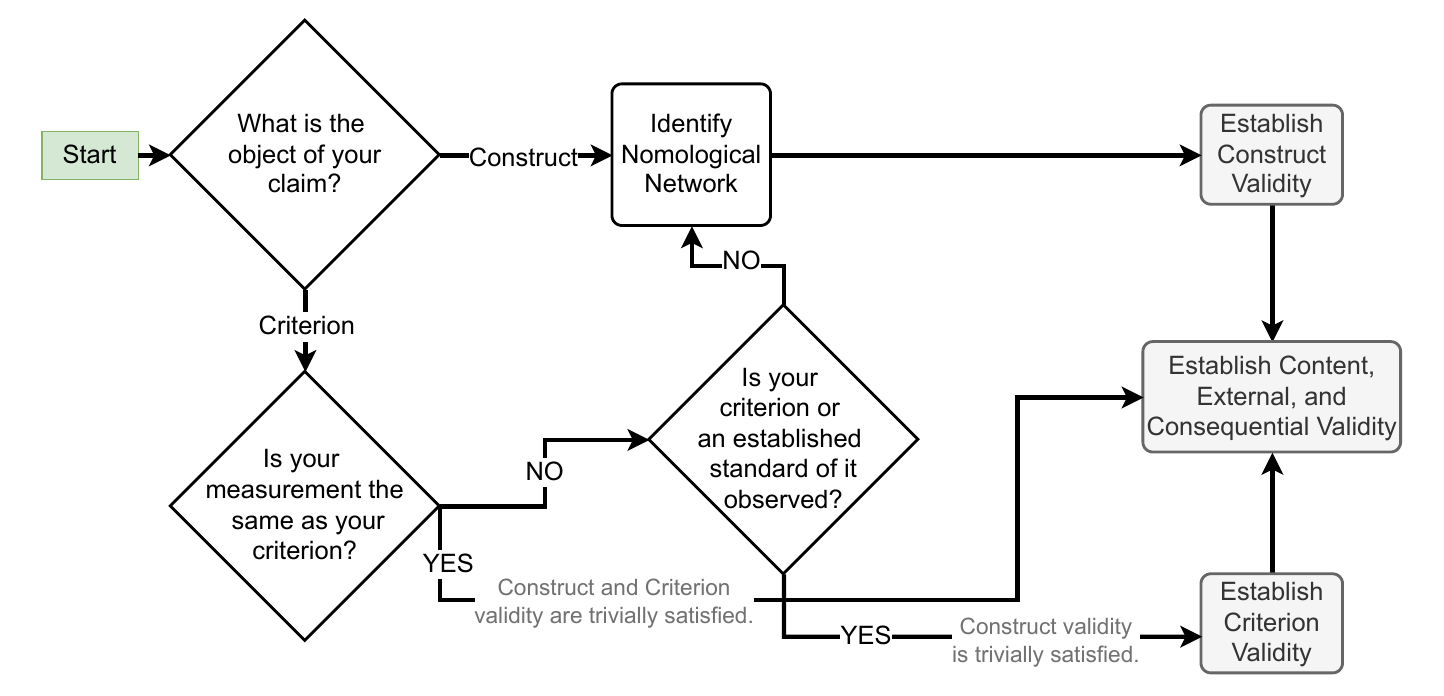}
    \caption{Decision process for establishing validity. For the decision processes that do not directly go through establishing construct or criterion validity, our argument is not that those forms of validity are irrelevant, but rather that they may be trivially satisfied in the context of the measurement, evaluation, and claim.}
    \label{fig:validity_decision_tree}
\end{figure}

\begin{figure}[t]
    \centering
    \includegraphics[width=\linewidth]{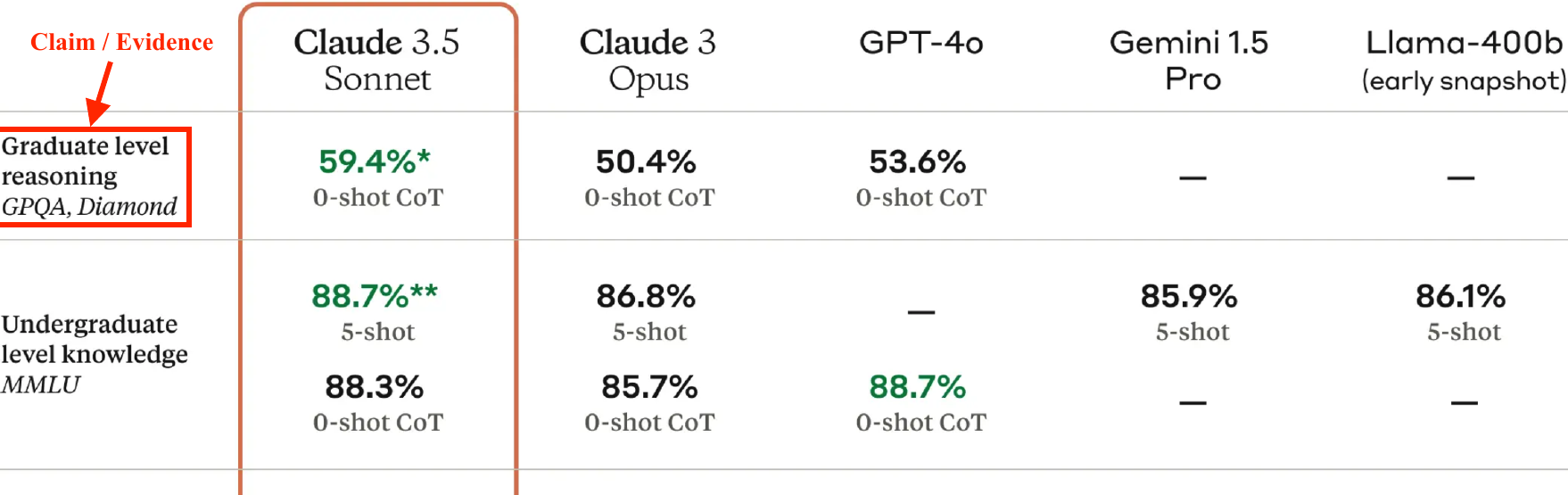}
    \caption{A real-world example of a developer presenting the Graduate-Level Google-Proof Q\&A Benchmark (GPQA) as a proxy for the graduate-level reasoning capabilities of their system (top left). Additionally, SWE-bench~\cite{jimenez2023swe} is used as a proxy for agentic coding, and $\tau$-bench~\citep{yao2024tau} as a proxy for agentic tool use~\citep{gpqaanthropic, Rein2023-so}.}
    \label{fig:gpqa_anthropic}
\end{figure}

\begin{table}[t]
    \centering
    \caption{A Graduate-Level Google-Proof Q\&A Benchmark (GPQA)~\citep{Rein2023-so} Application. A subjective score for validity—the standard for “reasonable” is demonstrating that obvious risks to invalidity are addressed: \okemoji: reasonable; \warnemoji: proceed with caution; \xemoji: insufficient. Even for a score of “reasonable,” there will be weaknesses in the evidence. The score is given because the strengths outweigh the weaknesses in terms of determining the validity of the claim from that evidence. This is never a binary classification nor complete, and should rather be a cyclic process—for instance, as our forms of what constitutes graduate-level chemistry may evolve over time and from school to school.}
    \label{tab:gpqa_report_card_}
    \resizebox{\textwidth}{!}{
    \begin{tabular}{|p{8.5cm}|c|c|c|c|c|}
\hline
\multicolumn{6}{|l|}{\textbf{Claims from Graduate-Level Google-Proof Question Answering (GPQA) Benchmark Accuracy Report Card}} \\
\hline
\textbf{Claims} & \textbf{Content} & \textbf{Criterion} & \textbf{Construct} & \textbf{External} & \textbf{Consequential} \\
\hline
1. AI systems can accurately answer \textit{graduate-level specialized multiple-choice questions} in biology, physics, and chemistry. 
    & \okemoji & \okemoji & \okemoji & \okemoji & \warnemoji \\
\hline
2. AI systems can accurately answer \textit{graduate-level specialized questions} in specialized scientific domains.
    & \warnemoji & \warnemoji & \warnemoji & \warnemoji & \warnemoji \\
\hline
3. AI systems can exhibit \textit{general reasoning abilities} that can transfer beyond current human specialization.
    & \warnemoji & \xemoji & \xemoji & \xemoji & \warnemoji \\
\hline
\end{tabular}
    }
\end{table}

In this section, we categorize when and how different forms of validity are most critical for supporting a claim with measurements and evaluation, Figure~\ref{fig:validity_decision_tree}. While we maintain that all forms of validity are always necessary, some may be trivially satisfied depending on the measurement, evaluation, and claim context. Rather than applying uniform scrutiny to all forms of validity, we account for context-dependent nuances that make certain forms particularly significant in some cases; this is distinct from previous work~\citep{Wallach2025-hx}.

Our framework, as described by Figure~\ref{fig:validity_decision_tree}, is similar to~\cite{lissitz2007suggested}, who challenge the dominant paradigm that all forms of validity are subsumed by construct validity, as advocated by~\citep{Messick1995-vu}. They argue that this approach can obscure critical distinctions between types of test use and the kinds of evidence needed to justify them. They argue that by folding all these aspects into a single, construct-validity-centric framework, the field risks undervaluing empirical utility and pragmatic decision-making goals. For example, if an IMO accuracy reliably predicts college linear algebra question-answering accuracy, it may be considered valid for such uses, even if the construct it measures (e.g., ``mathematical reasoning'' or ``pattern recognition'') is poorly defined or theoretically contested. \cite{lissitz2007suggested} call for a functional approach to validation, one that aligns the type of evidence required with the specific role the test plays. This perspective opens the door to broader and more flexible evaluation strategies, particularly in applied contexts where operational outcomes matter more than construct fidelity.

Specifically,~\cite{lissitz2007suggested} consider three common relevant activities, which are related to the structure of our framework.

\begin{itemize}
    \item \textbf{Utility Determination.} Validity should be determined by how useful a test is for its intended purpose. A valid test enables appropriate and effective decisions in real-world contexts (Sec.~\ref{sec:scen1}-\ref{sec:scen23}).
    \item \textbf{Theoretical Support.} A test is valid insofar as it supports the theory guiding its development or use (Sec.~\ref {sec:construct_object_claim}).
    \item \textbf{Impact Evaluation.} Validity must account for the outcomes of using a test, including whether the decisions it informs lead to beneficial, fair, or intended consequences (Sec.~\ref{sec:conseq_val}).
\end{itemize}

Our frameworks shares this perspective and develops a strategy to implement it for AI systems in the following.

Importantly, a claim can (and should~\citep{Wang2024-jz}) be supported by many evaluations and measurements. However, for simplicity and without loss of generality, we focus on a single measurement and evaluation.

Recall that the object of a claim can be a criterion (directly measurable) or a construct (abstract and not directly measurable). Our primary considerations for investigating and prioritizing forms of validity are determined by the following (Figure~\ref{fig:three_components}):

\begin{enumerate}
    \item Is the object of the claim a criterion (e.g., linear algebra textbook question accuracy) or a construct (e.g., reasoning ability)?
    \item Furthermore, is the measurement the same as the object of the claim (e.g., evaluating IMO accuracy when IMO accuracy is also the object of the claim)?
    \item Finally, if the measurement and object of the claim are different, does the measurement directly imply the claim, or does it require a mediating construct  (e.g., does IMO accuracy imply linear algebra textbook question accuracy, OR does it imply mathematical reasoning ability, which implies linear algebra textbook question accuracy)?
\end{enumerate}

The five forms of validity we foreground are relevant in different ways. Ideally, we should validate a claim by directly measuring the object it is about in the context in which we want to make the claim. However, this is often not possible. In any case, we must always validate content validity and external validity. Additionally, when we perform a measurement but the object of the claim is a different criterion (e.g., evaluate IMO problem solving $\rightarrow$ make claims about linear algebra textbook problem solving), criterion validity is most important. Criterion validity ensures that the measurement reliably predicts the object of the claim (predictive validity) or an established external standard of the object of the claim (concurrent validity). When neither the object of claim nor an established external standard is available, we may validate the claim through an intermediate construct (evaluate IMO problem solving $\rightarrow$ infer mathematical reasoning $\rightarrow$ make claims about linear algebra textbook problem solving), requiring construct validity of tests of mathematical reasoning with respect to the downstream use of linear algebra textbook problem solving.

When the object of the claim is itself a construct, and we directly measure and evaluate its proxies (e.g., evaluate IMO accuracy $\rightarrow$ make claims about mathematical reasoning), construct validity is essential to determine whether the measurement genuinely measures the intended construct rather than an unrelated or superficial correlation. This requires {\em structural validity} to ensure that the measurement reflects the internal structure expected of the construct (e.g., subskills of reasoning are coherently represented), {\em convergent validity} to demonstrate that the measurement correlates with other measures of the same or closely related constructs (e.g., aligns with expert assessments of reasoning), and {\em discriminant validity} to show that it does not correlate with measures of unrelated constructs (e.g., verbal fluency or memorization), guarding against misleading associations. This is also necessary when we aim to validate a claim about a construct with measurements and evaluations of proxies of other constructs (e.g., evaluate IMO problem solving $\rightarrow$ infer logical reasoning $\rightarrow$ make claims about mathematical reasoning). In this case, a Nomological Network—an understanding of the relationship between logical and mathematical reasoning and their observables—is paramount.

\begin{figure}
    \includegraphics[width=\linewidth]{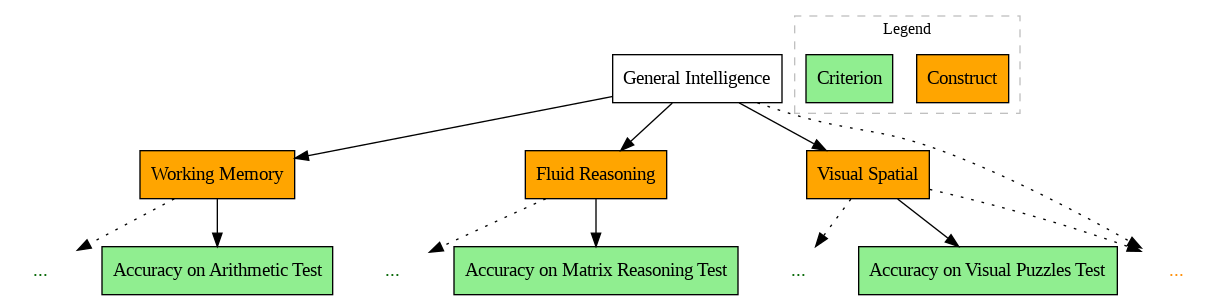}
    \caption{This figure illustrates a nomological network for human general intelligence according to the Wechsler Intelligence Scale, adapted from~\cite{Canivez2017-bs}. {\em We note that this breakdown of general intelligence is for illustrative purposes only; it does not necessarily translate to artificial general intelligence.} The network consists of background concepts (blue w/ rounded corners) linked by hypothesized associations, reflecting abstract expectations. Observable indicators (green w/ sharp corners) represent criteria, measurable variables used to assess the constructs. Establishing a robust nomological network is critical to ensuring construct validity by demonstrating abstract coherence, convergent validity, and discriminant validity within empirical research.}
    \label{fig:general_intelligence}
\end{figure}

{\em A claim about a construct cannot be validated in isolation—instead, it gains meaning and validity through its relationships with other constructs and observable measures through the scientific process.} Cronbach and Meehl’s nomological network~\citep{Cronbach1955-sc} provides a rigorous, albeit historically challenging to operationalize, way to reason about constructs within a broader abstract and empirical system, allowing for more scientifically robust validation. A nomological network is a conceptual framework that maps the relationships between constructs and criteria. Figure~\ref{fig:general_intelligence} gives an example from human psychology. An explicit nomological network for AI constructs, while vital, remains a missing piece in efforts toward valid AI evaluations, limiting the establishment of validity when constructs are involved. Although a detailed treatment of nomological networks is beyond the scope of this work, we emphasize their importance in establishing validity and explicitly indicate where they are necessary in our framework. We refer the reader to Cronbach and Meehl’s seminal work for more detail~\citep{Cronbach1955-sc}.

Next, we illustrate our framework for determining validity.

\section{Application of our Framework}
\label{sec:application}
We focus in the main text on GPQA (Graduate-Level Google-Proof Question Answering) accuracy as our measurement and evaluation. We then investigate claims of varying generality commonly made from this evaluation~\citep{Buntz2025-uc,Rein2023-so}. Additional examples are in Appendix~\ref{sec:casestudies}. This clarifies that a given measurement may not support broad claims, yet it can still be highly useful for supporting more narrowly defined ones. This adds necessary nuance to the discourse on validity in AI assessment.

Table~\ref{tab:gpqa_report_card_} summarizes the following example of applying our framework to assess the validity of claims from GPQA multiple-choice question-answering accuracy. We supplement this section with detailed case studies in the context of evaluating popular vision and/or language AI systems in Appendix~\ref{sec:casestudies}.

\paragraph{Sources of Validity Evidence.} While we restrict ourselves to the evidence of validity provided in the GPQA paper for the previous analysis for brevity and simplicity, establishing validity can (and should) be done across multiple asynchronous studies and various stakeholders.

Furthermore, while we can begin with a claim and use our framework to design and identify the necessary types of measurements and evaluations to support it, i.e., the scientific process; we can also do this in reverse. The latter is increasingly common in practice, given the exploratory nature of many of our studies of AI systems.~\cite{Adcock2001-lc}’s framework for measurement validity begins with the construct and reasons about developing valid measurements, as does~\cite{Wallach2025-hx}’s framework for evaluating generative AI. Recognizing the iterative nature of validation and that all measurements and evaluations are limited to some degree, our framework allows for the possibility of deriving valid claims even from imperfect measurements. Such claims are necessarily narrower in scope and generality than those supported by more complete or ideal measurements. We see this flexibility as a key practical advantage, particularly in modern contexts where the stakeholders conducting measurements and those making claims from evaluations may differ, and where some information needed to establish full validity may be unavailable.

We first consider the setting where the object of the claim is a criterion, and the desired claim in this setting looks like this: ``A higher measurement score predicts a higher/lower criterion score.'' ``A higher GPQA accuracy predicts a higher PhD qualifying exam accuracy.'' In this case, the measurement and the criterion object of the claim can be (1) identical (GPQA accuracy vs. new GPQA accuracy), (2) proxies of the same underlying construct (GPQA accuracy vs. general scientific question-answering accuracy), or (3) proxies of two different but related underlying constructs (GPQA accuracy (scientific reasoning) vs. medical diagnostic accuracy (medical reasoning)). (4) We then consider the setting where the object of the claim is a construct.

In each case, we must justify why the measurement supports the claim.

\subsection{Criterion-Aligned Evidence (1): The object of the claim is a criterion and is measured as evidence} \label{sec:scen1}
In scenario 1, we are primarily concerned with content and external validity, i.e., does our measurement cover the relevant content of the criterion, and does it generalize to relevant contexts beyond that of the measurement?  Construct validity and criterion are trivially satisfied as a consequence of directly measuring and evaluating the object of the claim. This could be because the hard work of systematization and operationalization of the construct we would have otherwise attempted to measure and evaluate has already been done~\citep{Adcock2001-lc,Wallach2025-hx}.

\paragraph{Identical Measurement and Criterion.} {\em Example}. Our object of claim is a criterion: multiple-choice questions accuracy in physics, chemistry, and biology\footnote{Thresholding is commonly used in real-world decision-making to transform continuous measurements (e.g., confidence scores) into binary or categorical outcomes (e.g., pass/fail, high risk/low risk). The choice of threshold can profoundly affect both evaluation and claim validity: even a perfectly measured property may lead to an invalid claim if the threshold does not align with the intended context or the actual consequences of misclassification. This is known in psychometrics as standard testing~\citep{cizek2007standard}.}. We then aim to support claims about an AI system’s accuracy on such questions by measuring and evaluating the system’s accuracy on the GPQA dataset; we must:

\begin{itemize}
    \item Establish content validity: GPQA has expert-curated questions, which enhance content validity by ensuring relevance and rigor across biology, physics, and chemistry, with the performance gap between experts and non-experts indicating effective assessment of specialized knowledge. However, the criteria may inadvertently exclude certain relevant topics, potentially skewing subfield representation. Systematic content mapping and expert diversity analysis can strengthen validity by ensuring comprehensive coverage and mitigating selection biases. In modern AI, red-teaming can also help identify the coverage gaps that hinder content validity~\citep{perez2022red}.
\end{itemize}

If content validity holds and one does not expect that the context of measurement is different than the context of the claim, then the claim can be supported by the measurement. However, if the claim must hold in a different context than the measurement, one must also:

\begin{itemize}
    \item Establish external validity: The GPQA measurement reflects real-world conditions, with human experts developing questions and a measurement format aligned with academic multiple-choice assessments, so the context of measurement is aligned with the context of the claim in this sense. However, the human assessment may not generalize beyond the measurement context, and without comparison to other multiple-choice science tests, its generalizability remains unverified. To strengthen external validity, validation against diverse question formats, question types, and other variations of context is necessary.
\end{itemize}

This setting is commensurate with traditional AI benchmarking practices. Many AI benchmarks have focused on these forms of generalization, including classical generalization~\citep{Vapnik1971-hp} and out-of-distribution generalization~\citep{Shimodaira2000-sl}. By ensuring strong content and external validity, such benchmarks provide a solid foundation for validating claims for known and directly measurable criteria.

\subsection{Criterion-Adjacent Evidence (2-3): The object of the claim is a criterion, but a different object is measured as evidence}\label{sec:scen23}
For 2-3, different criteria that are either proxies of the same or different but related mediating constructs. Ideally, we additionally directly establish criterion validity. That is, establish that the object that is measured is predictive of the desired criterion or an established standard. Then, the existence of these mediating constructs may inform how we establish criterion validity, but we do not need to reason about them directly to make valid claims.

{\em Example.} Our object of claim is general scientific question-answering, and we want to use GPQA accuracy as evidence. We still need to demonstrate that the measurement covers relevant content and generalizes to all the contexts we want the claim to hold (content and external validity). However,  we must additionally establish that the measurement of GPQA accuracy is predictive of the scientific question-answering criterion or a validated standard, i.e.:

\begin{itemize}
    \item Establish Criterion Validity. Human expert accuracy provides a strong external criterion and validated standard, supporting concurrent validity, while the AI-expert performance gap reinforces the benchmark’s credibility. However, there is no evidence of predictive validity, as accuracy has not been tested against future performance on specialized assessments, and concurrent validity remains incomplete without correlations to established external measures of expertise, such as standardized exams in other fields. To strengthen criterion validity, correlations should be established with real graduate program exams for concurrent validity, and predictive validity studies should track the system's downstream performance across scientific domains.
\end{itemize}

However, if criterion validity is implausible in this way, we may attempt to leverage our understanding of the underlying structure in constructs and their known mapping to observables, when it is available, to establish validity, i.e., a nomological network. Importantly, such a nomological network often does not exist in the current paradigm of AI assessment.

When a nomological network is unknown, establishing validity becomes significantly more difficult, as there is no agreed-upon basis for interpreting how abstract constructs like `scientific reasoning' map to measurable criteria (e.g., `GPQA accuracy'). In such cases, evaluations risk being narrow or misleading—a system might excel at scientific reasoning in physics yet still lack scientific reasoning in psychology, and evaluators could erroneously assume success in one facet implies the same in another. Without explicit connections between sub-constructs and corresponding measurements, conflicting results may emerge, and different assessments might rely on unfounded inferences about a system’s capabilities. This lack of structure not only obscures whether a measurement provides meaningful evidence for a given claim but also undermines the reliability of validity assessments, leaving practitioners vulnerable to inflated claims and misguided deployment decisions.

When such a network is available, to establish construct validity, we utilize its facets: structural, convergent, and discriminant validity:

\begin{itemize}
    \item Establish Construct Validity: For brevity, please refer to the subsequent discussion in section~\ref{sec:construct_object_claim} on when the object of the claim is a construct.
\end{itemize}

Importantly, when the measurement and criterion are proxies of different constructs, we must also validate relationships between constructs in addition to their relationships to observables. Doing this also requires knowledge of a nomological network. For example, in Figure~\ref{fig:general_intelligence}, the accuracy on the arithmetic test and accuracy on the matrix reasoning test must go between working memory and fluid reasoning.

{\em Example.} Suppose we want to make a claim about AI systems' reasoning about medical diagnosis based on their reasoning ability. To evaluate this, we must first define what reasoning entails. Suppose a model evaluator interprets reasoning as scientific reasoning and uses the GPQA benchmark to measure it. However, the object of the claim is most related to medical reasoning. At this stage, defining reasoning in this way is neither inherently valid nor invalid.

Now, suppose we want to claim that strong GPQA performance translates into accurate medical diagnosis; this requires several inferential steps. GPQA (insufficiently) assesses scientific reasoning, while medical diagnosis likely relies on medical reasoning, potentially a different subspace of reasoning, according to one's nomological network. Establishing structural validity requires examining whether GPQA captures the key components of general reasoning relevant to medical decision-making. Without showing that GPQA performance reflects the same underlying capabilities as medical reasoning, claims about AI outperforming physicians based on GPQA remain unverified. Establishing convergent validity through latent variable modeling and item-response theory can help demonstrate if the measurement captures variance in the latent subspace shared between the two constructs that determine the outcome criterion.

\subsection{Construct-Targeted Evidence (4): The object of the claim is a construct}\label{sec:construct_object_claim}
In many cases, we want to validate a claim about a construct by evaluating its proxies. This looks like: ``A higher measurement score implies a higher latent capability, e.g., GPQA accuracy to scientific reasoning.'' Then, construct validity is paramount.

{\em Example.} Suppose the object of the claim is general reasoning and we want to make a claim about a system’s general reasoning ability by measuring GPQA accuracy. Here, we must establish all five of our forms of validity, especially construct validity (recall it is composed of structural, convergent, and discriminant validity):

\begin{itemize}
    \item Establish Construct Validity: Performance on GPQA aligns with success in structured question-answering tasks, suggesting some reasoning component. However, structural validity is unclear, as the test may not sufficiently capture the rank of reasoning. Convergent validity is unverified since GPQA accuracy has not been correlated with other explicit reasoning assessments, e.g., interactive human evaluation. Discriminant validity is also uncertain, as it remains unclear whether GPQA measures genuine scientific reasoning or simply domain-specific knowledge and memorization. Comparing performance to humans with access to Google is an attempt to do this, though the time constraints on answering the question may add some construct-irrelevant hardness relative to human-AI comparisons. To address these concerns, methods like factor analysis~\citep{Kim1979-nq} should be conducted to distinguish reasoning from memorization, for instance, and performance should be validated against other dedicated reasoning assessments.
\end{itemize}

Additionally, content and external validity must be established to confirm that the essential aspects of the construct are accurately measured and that findings generalize to unmeasured components. Moreover, criterion validity, when a construct-relevant criterion or established standard is available, can support construct validity since well-designed measurements should reliably predict external outcomes related to the same construct.

\subsection{Consequential Validity} \label{sec:conseq_val}
Consequential validity examines whether the real-world outcomes of decisions based on an assessment align with its intended purpose. In the case of GPQA, if the benchmark effectively measures scientific reasoning, AI models that perform well on it could support decision-making in scientific research or education. However, there is a risk of overgeneralization—high GPQA accuracy might lead to misinterpreting AI as possessing broad reasoning abilities when it may only excel at structured multiple-choice problems. In this case, there could be harmful consequences like replacing human workers with ill-suited technology.

For strong consequential validity, GPQA measurement must align with the reasoning skills they intend to measure, ensuring AI performance is interpreted within its actual capabilities. Clear performance guidelines should distinguish validated reasoning abilities from speculative claims, preventing misapplications of AI in scientific decision-making.

\section{Who Should Care?}
We highlight key active stakeholders in the current AI ecosystem; this list is not meant to be exhaustive or elevate the importance of these stakeholders above others.
\begin{itemize}
    \item {\em Researchers.} Scientific claims about model capabilities must be grounded in evidence. Our framework helps researchers articulate what a given evaluation supports, distinguish between measurement and claim, and avoid overstating results, making their work more rigorous, interpretable, and useful to others.
    \item {\em Policy Makers.} As benchmarks increasingly guide regulatory decisions (e.g., under the EU AI Act), our framework provides a tool to assess whether the measurements being used actually support the claims that matter, like risk, safety, or societal impact. It helps ensure policy is evidence-aligned.
    \item {\em Corporations.} Product claims, deployment readiness, and resource allocation hinge on evaluation. Our framework helps internal teams diagnose whether current evaluations support the intended claims or identify what additional evidence is needed, reducing wasted effort and increasing confidence in deployment decisions.
    \item {\em Funders.} Grant agencies, philanthropic foundations, and investors must decide which projects merit scarce resources. By clarifying the link between what is measured and what is claimed, our framework lets funders vet proposals, set verifiable milestones, and track return on investment, ensuring their dollars advance research that delivers on its stated impact.
    \item {\em Civil Society.} As AI systems are deployed across sensitive domains, civil society organizations can use our framework to critically interrogate the basis of performance claims. It provides a structured way to ask: what exactly is being claimed, what is being measured, do they align, and how does misalignment affect their direct (personal use) and indirect (integration into critical infrastructure) day-to-day exposure?
\end{itemize}

\paragraph{Collective Accountability.} Validity cannot be outsourced. It depends on alignment between those who develop measurements, perform evaluations, make claims, act on those claims, and are affected by decisions based on them. However, in practice, these roles are often distributed across stakeholders with different incentives and timelines; researchers, industry teams, regulators, and civil society actors rarely operate in lockstep.

Our framework provides shared structure and vocabulary to bridge this gap. It enables researchers to design evaluations that are not only technically sound but claim-aware.


Critically, we emphasize the need for an iterative feedback loop. Evaluations are not one-shot exercises; neither are the claims derived from them. As systems are deployed and additional evidence becomes available, whether through failure cases, distributional shifts, or evolving standards, this evidence must feed back into how we measure, interpret, and assess. A claim that may be valid under one context or at one point in time may become invalid as the world changes or as new use cases emerge.

This feedback loop must also operate across stakeholders. Claims made by one group (e.g., developers asserting safety) can prompt others (e.g., independent researchers or regulators) to test, contest, or refine those claims using new measurements and evaluation strategies. Rather than expecting perfect measurement upfront, the goal is a responsive evaluation ecosystem, one that updates as systems, stakes, and societal expectations evolve.

Working together does not require perfect consensus, but it does require transparency and shared responsibility. Validity is not a checklist; it is a process. By coordinating around that process, stakeholders can better ensure that claims about AI capabilities are not just persuasive but also defensible, grounded, and worthy of trust.

\section{Decision Making with Validity in Mind}
Decision‑making that foregrounds validity is inherently context‑dependent; no single, one‑size-fits-all method will suffice. The same claim and evaluation can map to a `go' decision for one context and a `no-go' decision for another. Given a claim and evidence, we therefore recommend the following steps to help stakeholders systematically translate context‑specific validity evidence into defensible decisions. 

\begin{enumerate}
\item {\em Define Risk Tolerance.} Establish, up front, the maximum residual risk each stakeholder is willing to accept given the system’s context and potential impact.

\item {\em Define the Scope of the Claim.} Populate our validity framework with markers that indicate where evidence is weak, moderate, or strong; flag unresolved risks; and log newly surfaced unknowns.

\item {\em Weight by Harm Potential.} Calibrate evidence bar to the gravity of potential harm: \\high‑consequence claims (e.g., reasoning for medical triage) require stringent conditions and independent replication, whereas low‑consequence claims (e.g., reasoning for puzzle-solving games) can proceed under proportionally lighter evidence and safeguards.
 
\item {\em Collective Review.} Convene developers, domain experts, risk owners, external auditors, and other stakeholders for a decision‑focused session that examines the validity of the claim at hand.

\item {\em Document Decision Making and Share Publicly.}
Record both the decision and its rationale:  
(i) \okemoji{} risk tolerance fully satisfied;  
(ii) \warnemoji{} specific gaps remain but are bounded by mitigation plans and a re‑evaluation schedule;  
(iii) \xemoji{} evidence is inadequate, triggering further data collection or analysis.  
Archive the reasoning and assign clear owners for any follow‑up work.
\end{enumerate}

\section{Conclusion}
Historically, AI evaluation has centered on benchmarks, emphasizing narrow technical progress while neglecting the validity of broader claims. This was acceptable when generative AI was primarily a research endeavor with limited impact. Today, as general-purpose AI systems are deployed more widely, traditional evaluation practices fail to capture real-world utility and risk enabling premature claims about readiness and robustness. To address this, we introduce a claim-centered framework grounded in five key forms of validity—content, external, criterion, construct, and consequential. We outline how these forms support four common mechanisms for linking evidence to claims, identify risks to each form of validity, and provide practical strategies for mitigation. Unlike prior work, we also offer guidance on reasoning under imperfect measurement conditions.

A central challenge in AI evaluation is the conceptual gap between measured performance and actual capability. Our framework systematically bridges this gap, ensuring that assessments are rigorous, context-sensitive, and scientifically grounded. By explicitly mapping the relationship between measurements, evaluations, and the claims they are used to justify, the framework helps prevent overgeneralization and supports more accurate interpretations of AI performance.

We argue that AI evaluation must be claim-aware, evidence-driven, and methodologically sound. Implicit in any evaluation is a network of relationships between constructs, measurements, and criteria—a nomological network. Yet, the lack of explicit articulation of these networks has hindered the alignment between what we measure and what we aim to understand. Without this clarity, evaluations risk misrepresenting model capabilities, leading to flawed conclusions and inappropriate applications.

By adopting a principled approach to validity, we move beyond surface-level benchmarks toward more transparent and reliable assessments. This shift is critical for developing and deploying AI systems that are trustworthy and aligned with real-world needs.

This work offers a theoretical foundation for validity-centered AI evaluation, setting the stage for empirical investigation. By clarifying how measurements, evaluations, and claims interact—and by emphasizing the role of nomological networks, we present a flexible framework applicable across AI domains and tasks. We highlight existing tools for probing different forms of validity and point to future directions focused on operationalizing this framework in high-stakes settings. {\em Building robust nomological networks that map AI constructs to measurable variables will be key to ensuring that evaluations yield not only performance metrics but also meaningful insights into real-world utility and risk.}

\section*{Acknowledgments}
AA is funded by the NSF GRFP and the Knight-Hennessy Fellowship. SB is funded by the Stanford Graduate Fellowship (SGF). SK acknowledges support by NSF 2046795 and 2205329, the MacArthur Foundation, and Schmidt Sciences. We thank Florian~Dorner, Moritz~Hardt, Tom~S\"{u}hr, Sang~Truong, and Serena~Wang for detailed comments on early drafts.

\bibliographystyle{plainnat}
\bibliography{main}

\appendix
\newpage
\section*{Appendix Table of Contents}
\startcontents[appendices]

\printcontents[appendices]{}{1}{\setcounter{tocdepth}{2}}

\begin{landscape}
    \section{Evidence of Validity}\label{sec:evidence}
    \begin{longtable}{p{3cm} p{5cm} p{5cm} p{5cm}}
\caption{Common risks to validity, investigation tools, and evidence exemplars.
} \label{tab:validity}\\
\toprule
\textbf{Validity} & \textbf{Common risks} & \textbf{Investigation Tools} & \textbf{Evidence Exemplar} \\
\midrule
\endfirsthead
\multicolumn{4}{l}{\textit{Table \ref{tab:validity} continued from previous page}}\\
\toprule
\textbf{Validity} & \textbf{Common risks} & \textbf{Investigation Tools} & \textbf{Evidence Exemplar} \\
\midrule
\endhead
\midrule
\multicolumn{4}{r}{\textit{Table continues on the next page}}\\
\endfoot
\bottomrule
\endlastfoot

\textbf{\makecell[l]{Content\\Validity}} &
\begin{itemize}[label=$\Box$, topsep=0pt, partopsep=0pt]
  \item Coverage deficiency
  \item Construct irrelevance
  \item Imbalanced mixture of content
\end{itemize}
&
\begin{itemize}[label=$\Box$, topsep=0pt, partopsep=0pt]
  \item Expert review
  \item Red‐teaming / adversarially designed evaluations
  \item Synthetic data generation or edge cases
\end{itemize}
&
\begin{itemize}[label=$\Box$, topsep=0pt, partopsep=0pt]
  \item Documentation of how test items comprehensively cover the construct
  \item Explicit mapping of test content to abstract frameworks or industry standards
  \item Coverage analysis
\end{itemize}
\\

\textbf{\makecell[l]{Criterion\\Validity}} &
\textbf{Predictive and Concurrent Validity}
\begin{itemize}[label=$\Box$]
  \item Criterion contamination
  \item Criterion deficiency
  \item Restricted range
  \item Temporal/other shifts
\end{itemize}
&
\begin{itemize}[label=$\Box$, topsep=0pt, partopsep=0pt]
  \item Real‐world longitudinal studies
  \item Real‐world behavioral testing
  \item Scaling‐law predictive models
  \item Validated criterion studies
  \item Periodic post‐deployment testing
\end{itemize}
&
\begin{itemize}[label=$\Box$, topsep=0pt, partopsep=0pt]
  \item Correlation with an existing validated benchmark or gold standard
  \item Evidence that higher scores in evaluation metrics predict real‐world utility
\end{itemize}
\\

\multirow{3}{*}{\parbox{3cm}{\textbf{\makecell[l]{Construct\\Validity}}}} 
& 
\textbf{Structural:}
\begin{itemize}[label=$\Box$]
  \item Rank deficiency
  \item Poor factor structure
  \item Item interdependence
  \item Response format bias
  \item Complex measurement range
\end{itemize}
&
\begin{itemize}[label=$\Box$, topsep=0pt, partopsep=0pt]
  \item Theory building and hypothesis testing
  \item Factor modeling
  \item Studies of process
\end{itemize}
&
\begin{itemize}[label=$\Box$, topsep=0pt, partopsep=0pt]
  \item Observed changes in test performance under controlled conditions
  \item Item‐test correlations
  \item Emergent substructures in model behavior
\end{itemize}
\\

& 
\textbf{Convergent:}
\begin{itemize}[label=$\Box$]
  \item Irrelevant or weakly related evaluations
  \item High measurement error in scoring
  \item Restricted range (ceiling/floor effects)
  \item Confounding (e.g., memorization, format)
\end{itemize}
&
\begin{itemize}[label=$\Box$, topsep=0pt, partopsep=0pt]
  \item Benchmark suites for a construct (e.g., reasoning)
  \item Representation probing (e.g., causal mediation analysis of embeddings)
\end{itemize}
&
\begin{itemize}[label=$\Box$, topsep=0pt, partopsep=0pt]
  \item High correlation with other measures that assess the same construct
  \item Empirical clustering of model behaviors that align with constructs
\end{itemize}
\\

& 
\textbf{Discriminant:}
\begin{itemize}[label=$\Box$]
  \item Construct overlap
  \item Format‐induced correlations
\end{itemize}
&
\begin{itemize}[label=$\Box$, topsep=0pt, partopsep=0pt]
  \item Orthogonal datasets
  \item Decomposable metrics
\end{itemize}
&
\begin{itemize}[label=$\Box$, topsep=0pt, partopsep=0pt]
  \item Low or non‐significant correlation with measures of distinct constructs
  \item Evidence that evaluation does not overlap with unrelated dimensions
\end{itemize}
\\

\textbf{\makecell[l]{External\\Validity}} &
\begin{itemize}[label=$\Box$, topsep=0pt, partopsep=0pt]
  \item Sample bias
  \item Unrealistic testing conditions
  \item Temporal variability
  \item Interaction effects
  \item Experimenter effects
  \item Task‐specific bias
\end{itemize}
&
\begin{itemize}[label=$\Box$, topsep=0pt, partopsep=0pt]
  \item Red‐teaming
  \item Stress testing
  \item A/B testing
  \item Transfer testing
  \item Population‐stratified evaluations
\end{itemize}
&
\begin{itemize}[label=$\Box$, topsep=0pt, partopsep=0pt]
  \item Performance comparisons across different populations, environments, or settings
  \item Sensitivity analysis showing consistent performance under varying conditions
  \item Independent replication of results in different contexts or regions
\end{itemize}
\\

\textbf{\makecell[l]{Consequential\\Validity}} &
\begin{itemize}[label=$\Box$, topsep=0pt, partopsep=0pt]
  \item Bias / Fairness
  \item Adaptive overfitting
  \item Misuse of results
  \item Unintended incentives
  \item Policy and systematic consequences
  \item Temporal and other shift
\end{itemize}
&
\begin{itemize}[label=$\Box$, topsep=0pt, partopsep=0pt]
  \item Stakeholder interviews and feedback loops
  \item Societal impact audits
  \item Ethical stress testing
  \item Stakeholder feedback
\end{itemize}
&
\begin{itemize}[label=$\Box$, topsep=0pt, partopsep=0pt]
  \item Documented instances of evaluation‐driven improvements in safety, reliability, and fairness
  \item Impact studies
\end{itemize}
\\

\end{longtable}
\end{landscape}
\FloatBarrier
\section{Validity} \label{sec:validity}
Validity refers to the extent to which a test accurately measures what it is intended to measure. Validity has a rich history, originally developed in the context of drawing valid conclusions from tests, much like how we now aim to draw valid conclusions from AI evaluations. One of the earliest forms of validity is face validity, which refers to the extent to which a test appears to measure what it claims to, based on intuitive judgment. For instance, one may ask if symbolic regression from BigBench~\citep{Srivastava2022-dz} even appears to measure reasoning. However, relying on face validity alone can be misleading. As Charles Mosier~\citep{Mosier1947-mk} famously observed:

“This form [face validity] is also gratifying to the ego of the unwary test constructor. It implies that his knowledge and skill in the area of test construction are so great that he can unerringly design a test with the desired degree of effectiveness in predicting job success or in evaluating defined personality characteristics, and that he can do this so accurately that any further empirical verification is unnecessary. So strong is this ego complex that if statistical verification is sought and found lacking, the data represent something to be explained away by appeal to sampling errors or other convenient rationalization, rather than by scientific evidence which must be admitted into full consideration.”

A more structured form of validity emerged with content validity, which ensures that a test comprehensively covers all relevant aspects of the construct it aims to measure. For instance, one may ask if mathematical problem-solving benchmarks cover all relevant aspects of reasoning. Content validity is also typically assessed through expert judgment rather than statistical validation. Charles Lawshe~\citep{Lawshe1975-lh} later formalized this concept with the Content Validity Ratio (CVR), a method for quantifying expert agreement on test content.

Moving toward empirical rigor, predictive validity assesses a test’s ability to forecast an outcome of interest, typically a future outcome. This concept, introduced by Robert Thorndike in the mid-20th century during the rise of standardized testing, became central to fields like educational assessment, employment testing, and aptitude measurement~\citep{Thorndike1949-nv}. For example, the predictive validity of SAT scores for college GPA or cognitive ability tests for job performance has led to their widespread use for other outcomes~\citep{Kobrin2008-gf}. In the context of AI evaluation, one may ask ``Does accuracy on IMO benchmarks predict accuracy in textbook linear algebra questions?'' While predictive validity measures the correlation between a test and a future outcome, concurrent validity measures the correlation between a test and a validated standard applied at the same time under the same conditions. Predictive and concurrent validity make up criterion validity~\citep{aera2014standards}.

While criterion validity is useful for assessing direct correlations between tests and desired criteria, its limitations became apparent when evaluating abstract constructs, like psychological traits, rather than simple outcome-based predictions. In their seminal work on construct validity,~\citep{Cronbach1955-sc} highlighted these limitations. For example, while SAT scores may predict GPA, they may not reliably measure intelligence, as GPA is influenced by grading biases and other factors. Recognizing the risks of relying solely on criterion-based validity, Cronbach and Meehl introduced construct validity, which assesses the extent to which a test truly captures the theoretical construct it purports to measure.

Two key sources of evidence necessary for construct validity introduced by Campbell and Fiske (1959) are~\citep{Campbell1959-un}:

\begin{itemize}
    \item Convergent validity—the degree to which a test correlates with other measures of the same construct.
    \item Discriminant validity—the degree to which a test does not correlate with measures of unrelated constructs.
\end{itemize}

Implicitly, this framework also includes structural validity~\citep{Cronbach1955-sc,Messick1995-vu}, which examines whether a test's internal structure aligns with the theoretical construct it is designed to measure. This is often assessed using factor analysis or other dimensionality evaluations.

Cronbach and Meehl categorize validity into three primary forms:

\begin{enumerate}
    \item {\em Content validity}---ensuring a test comprehensively represents the concept it aims to measure.
    \item {\em Criterion validity}---evaluating how well a test correlates with external measures, which include predictive and concurrent validity. Concurrent validity refers to a test's agreement with a validated measure applied at the same time under the same conditions.
    \item {\em Construct validity}---assessing the theoretical alignment between a test and its intended construct.
\end{enumerate}

Beyond these core types, external validity refers to the extent to which a study’s findings can be generalized beyond its specific conditions. External validity examines whether results hold across different populations, settings, and time periods. Campbell and Stanley~\citep{Campbell2015-mt} were among the first to systematically define external validity, identifying factors like selection bias and situational specificity as risks to generalizability.

In response to Cronbach and Meehl’s framework, which emphasized the theoretical and statistical relationships between measures,~\citep{Messick1995-vu,Messick1998-nb} introduced consequential validity on the basis that validity is not just about measurement accuracy but also about the real-world impact of test interpretation and use. However, unlike~\citep{Messick1995-vu}, we do not unify all facets of validity under construct validity. We adopt the view of~\citep{lissitz2007suggested} where the use of a measurement determines what is necessary to support validity. Importantly, this may not require construct validity.

\cite{Borsboom2004-zt} offers a different view: validity is a property of the test itself, and a test is valid if and only if it measures the construct it purports to measure. In this view, questions of use or consequence are orthogonal to validity; what matters is whether the test causally reflects variation in the construct. This perspective draws a clear boundary between measurement and interpretation, placing the burden of validity squarely on the psychometric relationship between construct and test score. While theoretically clean, this stance omits considerations critical to our context, namely, how test outputs are used to make decisions. We, therefore, depart from Borsboom's definition, instead adopting a broader view in which validity also encompasses downstream consequences and use cases, particularly when evaluating AI systems deployed in high-stakes settings.

While these validity concepts were originally developed for psychological and educational testing, they provide a powerful lens for evaluating AI models. In the next section, we examine how these classical validity forms translate into the context of modern AI evaluation.

\section{The (Co)Evolution of evaluations and claims} \label{sec:evolution}
\subsection{Vision}
\begin{figure}[t]
    \centering
    \includegraphics[width=\linewidth]{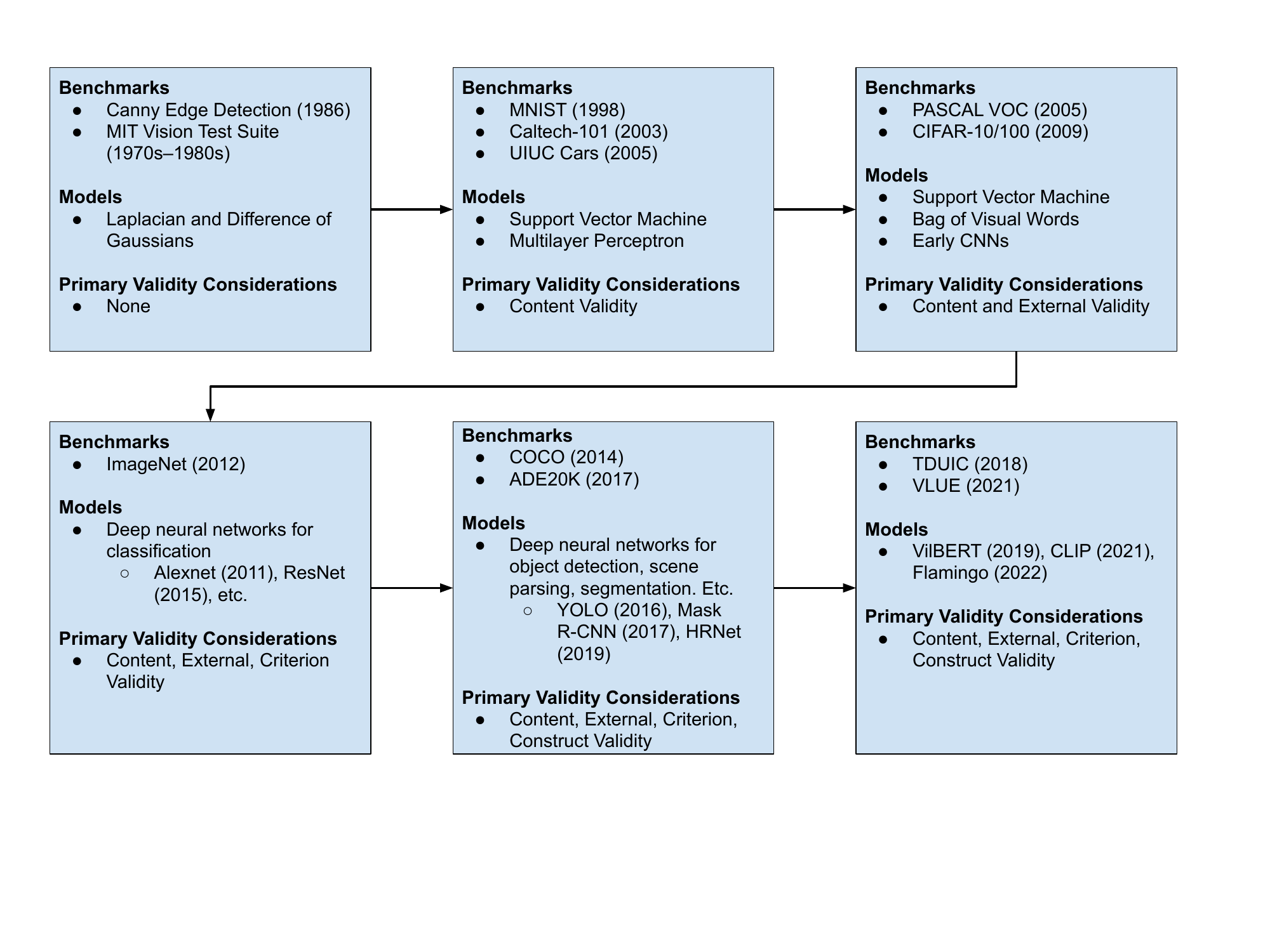}
    \caption{Coevolution of benchmarks, models, and the type of validity necessary for common conclusions for vision.}
    \label{fig:vision_coevolution}
\end{figure}

\begin{figure}[t]
    \centering
    \includegraphics[width=\linewidth]{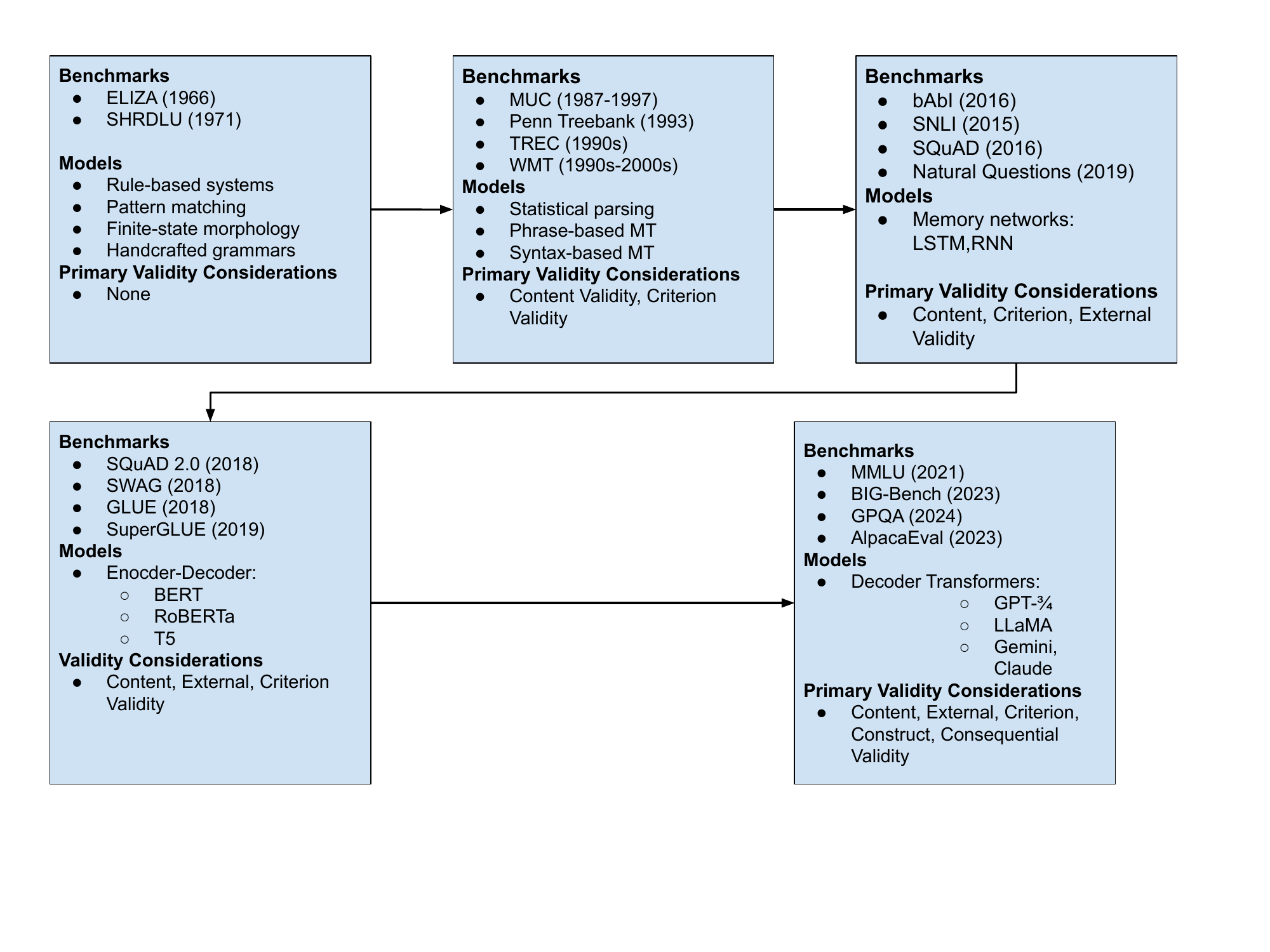}
    \caption{Coevolution of benchmarks, models, and the type of validity necessary for common conclusions for language.}
    \label{fig:language_coevolution}
\end{figure}
The evolution of AI benchmarks has been closely tied to the kinds of conclusions researchers aimed to draw and the evidence available at the time---Figure~\ref{fig:vision_coevolution}. In the 1960s to 1980s, benchmarks were hyper-localized, focusing on narrowly defined technical tasks like edge detection and simple shape recognition. The goal was primarily technical exploration—improving algorithmic efficiency—so the scope of conclusions was very narrow and directly supported by the evaluations carried out.

In the 1990s, AI benchmarks became more structured and began incorporating more applied tasks. A notable example is MNIST~\citep{Lecun1998-kd} for handwritten digit classification, which provided a standardized way to evaluate machine learning models. This trend continued into the early 2000s, with datasets such as UIUC Cars~\citep{Agarwal2002-lw} for vehicle detection and Caltech-101 (2003)~\citep{Fei-Fei2005-ge} for object recognition. While these benchmarks remained narrow in scope, they represented a step toward evaluating AI on more applied tasks, bridging the gap between theoretical research and practical applications. However, evaluations were still primarily designed for well-defined technical interests, with conclusions remaining local—focused on determining which techniques were most effective for the specific task being evaluated. During this period, researchers also became increasingly aware of content validity, recognizing that different datasets captured different aspects of classification tasks, which in turn influenced dataset design and evaluation methodologies~\citep{ponce2006dataset, torralba2011unbiased}.

By the mid-2000s, large-scale benchmarks such as PASCAL VOC (2007)~\citep{Everingham2010-qy} introduced greater complexity, expanding evaluation beyond simple classification tasks. Later, in the late 2000s, CIFAR-10 and CIFAR-100~\citep{Krizhevsky2009-tl} further pushed the field toward standardized comparisons in object recognition. During this period, criterion validity also gained prominence, as benchmark results were increasingly used to compare models in ways that suggested performance rankings carried external significance. However, construct validity remained largely unexplored—models were evaluated based on their outputs rather than on the reasoning processes behind their decisions. As a result, while evaluations became more sophisticated, they remained focused on performance metrics rather than deeper insights into model behavior. By this stage, the focus of AI evaluation began shifting from isolated dataset-specific improvements to broader claims about model robustness and transferability across different domains.


The 2010s marked a turning point with the ImageNet revolution. The introduction of ImageNet~\citep{Deng2009-xh} and the ILSVRC~\citep{Russakovsky2014-ka} competition (2010) provided large-scale, diverse, and complex benchmarks that dramatically reshaped AI research. During the early 2010s, the focus remained on improving accuracy in image classification and object detection. However, by the mid-2010s, AI evaluation expanded beyond leaderboards to real-world applications, particularly in medical imaging and autonomous driving. Researchers increasingly recognized the importance of content validity and external validity, leading to the widespread practice of testing models across multiple datasets to assess robustness.

As benchmark results gained influence, criterion validity became central—accuracy on ImageNet was frequently treated as a proxy for predicting downstream AI capabilities in vision. However, construct validity remained largely unaddressed in the early years. By the mid-2010s, early concerns emerged as researchers identified shortcut learning, adversarial vulnerabilities, and spurious correlations, leading to growing interest in understanding how models made decisions beyond raw accuracy. The rise of segmentation (COCO~\citep{Lin2014-pf}, ADE20K~\citep{Zhou2016-qv}) and video analysis benchmarks (Kinetics, AVA) reflected an effort to capture more complex real-world tasks, but fundamental concerns about model robustness and bias persisted.

In the 2020s, the rise of multimodal and foundation models introduced even greater evaluation challenges. Benchmarks such as VQA~\citep{Agrawal2015-if}, VLUE~\citep{Zhou2022-zq}, and TDIUC~\citep{Kafle2017-wl} attempted to assess multimodal reasoning, but defining what these benchmarks truly measured became increasingly difficult. Construct validity became a major concern as researchers debated whether these benchmarks genuinely assessed constructs like reasoning and understanding or merely exposed a model’s ability to exploit statistical correlations in large datasets (Sec.~\ref{sec:gaps}). Unlike earlier benchmarks, which primarily focused on accuracy, modern benchmarks aim to evaluate the latent properties of AI systems~\cite{gpqaanthropic}. However, fundamental questions about the validity of these evaluations remain unresolved, particularly in assessing generalization, robustness, and true reasoning ability.

Across these decades, benchmarks evolved alongside the conclusions stakeholders sought to make. Early benchmarks required little discussion of validity because they were purely technical exercises. As AI models became more ambitious and claims about their capabilities expanded, benchmarks had to keep up—introducing concerns about content, external, and criterion validity. More recently, as AI systems move toward multimodal reasoning and foundation models, discussions of construct validity have become central. As models grow in complexity, the challenge is no longer just about designing better benchmarks—it’s about defining what those benchmarks are actually supposed to measure in the first place.

\subsection{Language}
Language model benchmarks have seen an evolution from focusing on primarily basic questions of criterion validity against human performance to more nuanced considerations of other validity in more recent years---Figure~\ref{fig:language_coevolution}.
In the Blocks World Era (1960s-1980s), NLP evaluation was primarily qualitative and demonstration-based, lacking standardized metrics entirely. Systems like ELIZA (1966)~\citep{weizenbaum1966eliza} and SHRDLU (1971)~\citep{winograd1972understanding} were evaluated through anecdotal observations of how users interacted with them in highly constrained environments. ELIZA simulated a psychotherapist using simple pattern matching, while SHRDLU operated in a ``blocks world'' where users could issue commands to manipulate virtual objects. Validity considerations during this era were minimal and largely implicit. Content validity was severely limited by extremely narrow domains, criterion validity was nonexistent without standardized measurements, and construct validity wasn't addressed as researchers weren't attempting to measure specific capabilities like ``reasoning'' or ``understanding.'' External validity was particularly weak as systems couldn't generalize beyond their constrained environments. Success was measured simply by the system's ability to maintain seemingly intelligent conversations or follow instructions rather than through quantitative performance metrics or validity criteria.
The North Star Era (1990s-2000s) marked a paradigm shift toward empirical evaluation with standardized benchmarks inspired by information retrieval traditions, where benchmarks with quantitative metrics and clearly defined train, validation ,and test split gave the field a proverbial ``North Star'' to aim towards. Initiatives like the Message Understanding Conferences (MUC) and the Penn Treebank established common datasets, clearly defined tasks, and metrics such as precision, recall, and F-score for comparing systems. This era introduced the first rigorous validity considerations, though still narrow in scope. Benchmarks like TREC~\citep{harman1993overview} and WMT~\citep{bojar2017findings} established improved criterion validity through standardized metrics that allowed consistent measurement across systems and time. Content validity improved but remained limited to specific linguistic tasks. Nascent construct validity concerns emerged as researchers began considering what abilities their tasks were actually measuring. However, external validity remained largely unaddressed as benchmarks weren't designed to generalize beyond their specific contexts. Consequential validity still wasn't a major consideration, as NLP applications weren't yet widely deployed with significant societal impact.

In the early 2010s, many language benchmarks, such as SQuAD~\citep{Rajpurkar2016-il} and SNLI~\citep{Bowman2015-pr}, focused on individual tasks such as reading comprehension or natural language claims such as entailment or contradiction. The primary focus was on establishing baseline comparisons against human performance to create criterion validity for the benchmarks. However, such benchmarks had limitations to other aspect,s such as content validity due to limited focus on specific linguistic tasks and face validity due to narrow objectives and methods used to solve the task (both SQuAD and SNLI can be cast as relatively simple classification problems for which we can measure a gold standard of correctness). Other validity types were not heavily considered at this time.

In the mid to late 2010s, the field began to focus more on multi-task evaluation, which was represented by benchmarks such as GLUE~\citep{Wang2018-fk} and SentEval~\citep{Conneau2018-db}. During this time, emerging validity concerns became prominent. More sophisticated human baselines were required to maintain criterion validity,and  broader task coverage led to great content validity. However, concerns about the underlying mechanisms that could explain performance began to emerge, which reflects early concerns about construct validity. 

In the late 2010s there were key changes in language model evaluation. Benchmarks like SuperGLUE~\citep{Wang2019-th} aimed to resolve validity concerns with rigorous multi-annotator baselines, broader task selection, more attention to the demographics of annotators, and the first considerations of social impact and gaming. However, the lack of structural validity evidence and external validation remained as challenges. There were also few analyses of convergent/discriminant validity in studies. 

The 2020s marked a shift toward comprehensive knowledge evaluation with benchmarks like MMLU \citep{Hendrycks2020-tn}, reflecting a growing recognition that language models were advancing beyond narrow linguistic tasks to broader knowledge and reasoning capabilities. MMLU introduced several innovations in validity considerations: it established expert-level performance as the criterion validity benchmark rather than average human performance, expanded content validity through coverage of 57 subjects across multiple domains, and highlighted crucial external validity concerns through studies showing sensitivity to answer ordering and other conditions that should not have an effect on the downstream performance for an “intelligent” agent (as measured with respect to an expert).
The evolution of MMLU reflects broader trends in the field's approach to validity. Earlier benchmarks like SQuAD primarily focused on criterion validity through human performance comparisons, while MMLU attempted to address multiple validity types simultaneously. However, new challenges emerged: convergent validity became more complex as models showed inconsistent performance across related tasks (e.g., philosophy versus morality questions), and discriminant validity concerns arose around distinguishing between memorization and reasoning capabilities.
This progression has led to the current state of language model evaluation, characterized by greater sophistication in validity considerations but also a clearer recognition of inherent limitations. Recent work has highlighted the need for better convergent validity across benchmarks and more robust methods for assessing reasoning abilities. The field has moved from treating benchmarks as simple performance metrics to viewing them as complex instruments requiring multiple types of validation evidence~\citep{Ruan2024-fw}.

\newpage

\section{Case Studies}\label{sec:casestudies}

\subsection{GPQA}
\FloatBarrier
\begin{table}[h]
    \centering
    \caption{A Graduate-Level Google-Proof Question Answering Benchmark (GPQA)~\citep{Rein2023-so} Application. A subjective score for validity—the standard for “reasonable” is demonstrating that obvious risks to invalidity are addressed: \okemoji: reasonable; \warnemoji: proceed with caution; \xemoji: insufficient. Even for a score of “reasonable,” there will be weaknesses in the evidence. The score is given because the strengths outweigh the weaknesses in terms of determining the validity of the claim from that evidence. This is never a binary classification nor complete, and should rather be a cyclic process---for instance, as our forms of what constitutes graduate-level chemistry may evolve over time and from school to school.}
    \label{tab:gpqa_report_card}
    \resizebox{\textwidth}{!}{
    \begin{tabular}{|p{8.5cm}|c|c|c|c|c|}
\hline
\multicolumn{6}{|l|}{\textbf{Claims from Graduate-Level Google-Proof Question Answering (GPQA) Benchmark Accuracy Report Card}} \\
\hline
\textbf{Claims} & \textbf{Content} & \textbf{Criterion} & \textbf{Construct} & \textbf{External} & \textbf{Consequential} \\
\hline
1. AI systems can accurately answer \textit{graduate-level specialized multiple-choice questions} in biology, physics, and chemistry. 
    & \okemoji & \okemoji & \okemoji & \okemoji & \warnemoji \\
\hline
2. AI systems can accurately answer \textit{graduate-level specialized questions} in specialized scientific domains.
    & \warnemoji & \warnemoji & \warnemoji & \warnemoji & \warnemoji \\
\hline
3. AI systems can exhibit \textit{general reasoning abilities} that can transfer beyond current human specialization.
    & \warnemoji & \xemoji & \xemoji & \xemoji & \warnemoji \\
\hline
\end{tabular}
    }
\end{table}
\FloatBarrier
\paragraph{Description of dataset.} The GPQA (Graduate-Level Google-Proof Question Answering) benchmark is a challenging dataset comprising 448 multiple-choice questions crafted by domain experts in biology, physics, and chemistry~\citep{Rein2023-so}. These questions are designed to be exceptionally difficult, with experts holding or pursuing PhDs in the respective fields achieving an accuracy of 65\% (74\% when excluding clear mistakes identified retrospectively). Notably, highly skilled non-expert validators, even with unrestricted web access and spending over 30 minutes per question, attained only 34\% accuracy, underscoring the "Google-proof" nature of the dataset. State-of-the-art AI systems also find this benchmark challenging; for instance, a GPT-4 based model achieved 39\% accuracy. The GPQA dataset serves as a valuable resource for developing scalable oversight methods, aiming to enable human experts to effectively supervise and extract truthful information from AI systems that may surpass human capabilities.

\textbf{Object of Claim: } Multiple-choice questions in biology, physics, and chemistry accuracy.\\
\textbf{Claim 1:} AI models can accurately answer graduate-level specialized multiple-choice questions in biology, physics, and chemistry — criterion is accuracy on such questions.\\
\textbf{Evidence:} Accuracy on multiple-choice questions in biology, physics, and chemistry.

\textbf{Validity of Claim from Evidence:}
\begin{enumerate}
    \item \textbf{Content Validity \okemoji}
    \begin{itemize}
        \item \textit{Strength:} Expert-curated questions ensure high-quality, relevant content across key topics in biology, physics, and chemistry. The performance gap between experts and non-experts confirms the questions assess specialized knowledge.
        \item \textit{Weakness:} The dataset's construction criteria may exclude some relevant questions, potentially leading to over- or underrepresentation of certain subfields.
        \item \textit{Suggestions:} Conduct systematic content mapping across subfields to ensure balanced representation. Include expert diversity analysis to mitigate potential biases in question selection.
    \end{itemize}
    \item \textbf{Criterion Validity \okemoji}
    \begin{itemize}
        \item \textit{Strength:} Human expert accuracy provides a meaningful external criterion, reinforcing concurrent validity.
        \item \textit{Weakness:} Criterion validity could be stronger with comparisons to other specialized science Q/A benchmarks. Predictive validity is untested—no evidence that GPQA accuracy predicts future performance on exams or coursework, for example.
        \item \textit{Suggestions:} Compare performance with established science Q\&A benchmarks. Conduct longitudinal studies tracking how benchmark performance predicts success on real graduate exams.
    \end{itemize}
    \item \textbf{Construct Validity \okemoji}
    \begin{itemize}
        \item Since the claim is strictly about accuracy on a defined criterion, construct validity is not necessary to evaluate this specific claim.
    \end{itemize}
    \item \textbf{External Validity \okemoji}
    \begin{itemize}
        \item \textit{Strength:} The test mirrors a real-world setting—human experts develop the questions, and the evaluation format aligns with academic multiple-choice assessments. GPQA includes diverse topics within its disciplines.
        \item \textit{Weakness:} Similar to the criterion validity gap, GPQA accuracy is not compared to other multiple-choice science tests, leaving external generalization unverified.
        \item \textit{Suggestions:} Validate against different question formats and compare performance across multiple science benchmarks.
    \end{itemize}
    \item \textbf{Consequential Validity \warnemoji}
    \begin{itemize}
        \item \textit{Strength:} The AI-expert performance gap prevents premature claims of AI superiority, mitigating risks of overestimating AI scientific knowledge. However, models have quickly improved in this benchmark\footnote{\href{https://www.youtube.com/watch?v=ZANbujPTvOY}{https://www.youtube.com/watch?v=ZANbujPTvOY}.}. GPQA-trained models could support science education as study tools.
        \item \textit{Weakness:} If AI models reach high accuracy, stakeholders may overgeneralize their competence, assuming they have true expertise in physics, biology, and chemistry, despite lacking deeper scientific reasoning skills.
        \item \textit{Suggestions:} Develop clear guidance for stakeholders on interpreting results. Create documentation explicitly distinguishing multiple-choice performance from broader scientific expertise.
    \end{itemize}
\end{enumerate}

\textbf{Object of Claim:} Domain-specific scientific competency.\\ 
\textbf{Claim 2:} AI models can accurately answer graduate-level questions in specialized scientific domains—criterion is accuracy on such questions.\\
\textbf{Evidence:} Accuracy on [N] multiple-choice questions in biology, physics, and chemistry.

\textbf{Validity of Claim from Evidence:}
\begin{enumerate}
    \item \textbf{Content Validity \warnemoji}
    \begin{itemize}
        \item \textit{Strength:} Expert-curated, high-quality questions covering key topics in biology, physics, and chemistry. Non-expert performance gap supports specialization.
        \item \textit{Weakness:} Limited to three disciplines, excluding other specialized scientific domains (e.g., medicine, engineering). Only Q/A questions, excluding fill-in-the-blank or open-ended questions.
        \item \textit{Suggestions:} Expand questions to include other scientific subdomains. Conduct systematic content mapping across subfields to ensure balanced representation. Include expert diversity analysis to mitigate potential biases in question selection. 
    \end{itemize}
    \item \textbf{Criterion Validity \warnemoji}
    \begin{itemize}
        \item \textit{Strength:} Human expert accuracy serves as a strong external criterion (concurrent validity). AI-expert performance gap reinforces benchmark credibility.
        \item \textit{Weakness:} No predictive validity—GPQA accuracy is not tested against future performance on other specialized assessments.
        \item \textit{Suggestions:} Establish correlations with performance on real graduate program assessments. Develop predictive validity studies tracking model performance across time and domains.
    \end{itemize}
    \item \textbf{Construct Validity \warnemoji} {\em (importantly, this may be trivially satisfied if we have strong enough criterion validity.)}
    \begin{itemize}
        \item \textit{Strength:} Expert-curated questions in biology, physics, and chemistry are designed to capture fundamental aspects of specialized scientific knowledge. This suggests that the construct measured—domain-specific scientific competence—has meaningful representation, and high accuracy should correlate with understanding key scientific principles.
        \item \textit{Weakness:} GPQA’s focus on biology, physics, and chemistry limits its ability to capture the overall construct of “specialized scientific knowledge,” as other fields like medicine and engineering require different reasoning and knowledge structures. Moreover, the paper does not provide evidence linking GPQA performance to external measures of scientific competence (such as standardized test scores), leaving its alignment with related constructs unclear. Finally, the multiple-choice format may favor recognition or memorization over deeper analytical reasoning, potentially failing to capture key facets like synthesis and in-depth understanding.
        \item \textit{Suggestions:} To improve construct validity, expand GPQA to include additional domains (e.g., medicine, engineering) and correlate its scores with independent standardized assessments to establish convergent and discriminant validity. Additionally, incorporating alternative formats like open-ended questions and problem-solving tasks will better capture domain-specific scientific competence.
    \end{itemize}
    \item \textbf{External Validity \warnemoji}
    \begin{itemize}
        \item \textit{Strength:} Real-world, expert-created multiple-choice questions ensure relevance. Coverage across multiple subfields increases generalization within biology, physics, and chemistry.
        \item \textit{Weakness:} No evidence of generalization to other science assessments (e.g., (non-)multiple choice PhD qualifying exams).
        \item \textit{Suggestions:} Test generalization to other assessment formats including written exams, oral defenses, and research proposal evaluations.
    \end{itemize}
    \item \textbf{Consequential Validity \warnemoji}
    \begin{itemize}
        \item \textit{Strength:} AI-expert performance gap prevents overstating AI’s scientific capabilities; models could support science education.
        \item \textit{Weakness:} Risk of overgeneralization—high scores may be misinterpreted as broad scientific expertise beyond tested domains.
        \item \textit{Suggestions:} Create clear limitations documentation highlighting specific domains where evidence supports or doesn't support performance claims.
    \end{itemize}
\end{enumerate}

\textbf{Object of Claim:} Reasoning.\\
\textbf{Claim 3:} AI models exhibit general reasoning abilities.\\
\textbf{Evidence:} Accuracy on [N] multiple-choice questions in biology, physics, and chemistry.

\textbf{Validity of Claim from Evidence:}
\begin{enumerate}
    \item \textbf{Content Validity \warnemoji}
    \begin{itemize}
        \item \textit{Strength:} Covers multiple scientific disciplines, requiring some level of reasoning beyond factual recall.
        \item \textit{Weakness:} Multiple-choice format limits assessment of forms of reasoning like logical deduction, or abstract problem-solving.
        \item \textit{Suggestions:} Develop specific reasoning-focused questions that isolate logical deduction from domain knowledge. Include diverse reasoning types (inductive, deductive, abductive).
    \end{itemize}
    \item \textbf{Criterion Validity \xemoji}
    \begin{itemize}
        \item \textit{Strength:} Human expert accuracy serves as a real-world external criterion, and the AI-expert performance gap indicates a meaningful benchmark for reasoning capabilities.
        \item \textit{Weakness:} GPQA tests factual and applied knowledge rather than abstract reasoning skills. No predictive validity—performance on GPQA is not tested against other established reasoning benchmarks (e.g., LSAT-style logical reasoning or problem-solving tests).
        \item \textit{Suggestions:} Compare performance against established reasoning benchmarks like LSAT, GRE analytical, and domain-independent logical reasoning tests.
    \end{itemize}
    \item \textbf{Construct Validity \xemoji}
    \begin{itemize}
        \item \textit{Strength:} AI performance on GPQA correlates with success in structured question-answering tasks, suggesting some reasoning component. Additionally, the dataset can distinguish between human experts and non-experts.
        \item \textit{Weakness:} Does not separate reasoning from memorization—AI models may exploit dataset patterns rather than apply logical deduction. While non-experts with access to Google perform worse than experts, non-experts are given a limited time per question, which may not sufficiently show that models have not been trained on such questions. No convergent validity—GPQA accuracy is not correlated with performance on explicit reasoning assessments. No discriminant validity—It is unclear whether GPQA measures reasoning ability or just domain-specific knowledge.
        \item \textit{Suggestions:} Conduct factor analysis to distinguish reasoning from memorization. Demonstrate convergent validity with dedicated reasoning assessments and discriminant validity from pure knowledge recall.
    \end{itemize}
    \item \textbf{External Validity \xemoji}
    \begin{itemize}
        \item \textit{Strength:} GPQA questions require problem-solving across multiple disciplines, increasing the likelihood that some reasoning ability is being tested.
        \item \textit{Weakness:} Reasoning should generalize across domains, but GPQA only includes three scientific fields. No evidence that AI models with high GPQA accuracy perform well on general reasoning tasks outside science (e.g., logical puzzles, mathematical proofs, legal or philosophical reasoning).
        \item \textit{Suggestions:} Test performance on reasoning tasks across non-scientific domains including logic puzzles, mathematical proofs, and philosophical arguments.
    \end{itemize}
    \item \textbf{Consequential Validity \warnemoji}
    \begin{itemize}
        \item \textit{Strength:} If GPQA successfully measures reasoning, AI models excelling on it could serve as decision-support tools in scientific research or education.
        \item \textit{Weakness:} Overgeneralization risk—high GPQA accuracy may lead to misinterpreting AI as possessing broad, human-like reasoning abilities when it may only excel at structured multiple-choice problems.
        \item \textit{Suggestions:} Develop clear performance interpretation guidelines specifying which reasoning capabilities are supported by evidence versus which remain speculative.
    \end{itemize}
\end{enumerate}
\pagebreak
\subsection{ImageNet}
\FloatBarrier
\begin{table}[H]
    \centering
    \caption{An ImageNet~\citep{Deng2009-xh, Russakovsky2014-ka} Application. A subjective score for validity—the standard for “reasonable” is demonstrating that obvious risks to invalidity are addressed: \okemoji: reasonable; \warnemoji: proceed with caution; \xemoji: insufficient. Even for a score of “reasonable,” there will be weaknesses in the evidence. The score is given because the strengths outweigh the weaknesses in determining the validity of the claim from that evidence. This evaluation is an iterative process, acknowledging that both the benchmark and its interpretations may evolve over time.}
    \label{tab:imagenet_report_card}
    \resizebox{\textwidth}{!}{
        \begin{tabular}{|p{8.5cm}|c|c|c|c|c|}
    \hline
    \multicolumn{6}{|l|}{\textbf{Claims from ImageNet Validity Assessment Report Card}} \\
    \hline
    \textbf{Claims} & \textbf{Content} & \textbf{Criterion} & \textbf{Construct} & \textbf{External} & \textbf{Consequential} \\
    \hline
    1. ImageNet tests how well models learn complex associations between images and labels. & \okemoji & \okemoji & \okemoji & \okemoji & \warnemoji \\
    \hline
    2. ImageNet gauges the ability to learn semantically general visual features for object classification. & \warnemoji & \okemoji & \warnemoji & \warnemoji & \warnemoji \\
    \hline
    3. ImageNet measures overall visual understanding of a model. & \xemoji & \xemoji & \xemoji & \xemoji & \xemoji \\
    \hline
    \end{tabular}}
\end{table}
\FloatBarrier

\paragraph{Description of dataset.}  
ImageNet~\citep{Deng2009-xh, Russakovsky2014-ka} (specifically ILSVRC 2012) is a benchmark for predicting an image's label from a fixed set of 1000 diverse categories. The dataset—curated primarily from Flickr with human annotation—is evaluated using accuracy/error rate and precision/recall metrics.

\bigskip

\textbf{Object of Claim:} Predictive accuracy.\\
\textbf{Claim 1:} Model architectures can learn to accurately predict predefined image labels.\\
\textbf{Evidence:} Performance on accuracy/error rate and precision/recall metrics.

\textbf{Validity of Claim 1 from Evidence:}
\begin{enumerate}
    \item \textbf{Content Validity \okemoji}
    \begin{itemize}
        \item \textit{Strength:} The dataset covers 1000 diverse categories with extensive natural variability—including differences in poses, lighting, backgrounds, and fine-grained distinctions (e.g., different dog breeds)—making it well-suited to assess image–label associations.
        \item \textit{Weakness:} It is confined to static, natural RGB images and does not include other modalities (e.g., grayscale medical images or hyperspectral data) or dynamic contextual information (e.g., actions or inter-object relationships). Label noise may also affect accuracy metrics~\citep{northcutt2021pervasive}.
        \item \textit{Suggestions:} Clearly specify that ImageNet targets static natural images, and consider integrating supplementary datasets to represent additional image types or contextual settings.
    \end{itemize}
    \item \textbf{Criterion Validity \okemoji}
    \begin{itemize}
        \item \textit{Strength:} There is robust evidence that performance on ImageNet is both predictive of downstream task success (models excelling on ImageNet often perform well on benchmarks such as CIFAR or Caltech, and in real-world applications like wildlife classification~\citep{Norouzzadeh2018-yr}) and concurrent with human-annotated labels under similar conditions~\citep{he2020momentum, zhai2022scaling, Kornblith2018-ho}.
    \end{itemize}
    \item \textbf{External Validity \okemoji}
    \begin{itemize}
        \item \textit{Strength:} The dataset is representative of real-world natural images, and its utility has been demonstrated under varying conditions (differences in image quality, size, and even in applications to non-traditional domains such as medical imaging~\citep{Irvin2019-lj} and adversarially constructed settings~\citep{Salaudeen2024-yu}. Note, this is not about trained model performance (e.g.,~\cite{Recht2019-at}); it is about the external validity of model ability to learn and predict accurately, i.e., necessitates training and evaluating in a new setting rather than transporting trained models to a new setting.
    \end{itemize}
    \item \textbf{Construct Validity \okemoji}
    \begin{itemize}
        \item Since the claim is strictly about accuracy on a defined criterion, construct validity is not necessary to evaluate this specific claim.
    \end{itemize}
    \item \textbf{Consequential Validity \warnemoji}
    \begin{itemize}
        \item \textit{Strength:} The clear quantification of labeling accuracy offers a concrete performance metric, facilitating transparent and reproducible comparisons.
        \item \textit{Weakness:} There is a risk that high ImageNet accuracy may be misinterpreted as reflecting comprehensive visual understanding, potentially leading to overconfident real-world deployments.
        \item \textit{Suggestions:} Advise stakeholders that ImageNet performance should be interpreted strictly as a measure of static image classification and that complementary evaluations are necessary to assess broader aspects of visual intelligence.
    \end{itemize}
\end{enumerate}

\bigskip

\textbf{Object of Claim:} Learning of semantically general visual features.\\
\textbf{Claim 2:} ImageNet evaluates the ability of models to learn transferable visual features that are useful for object classification.\\
\textbf{Evidence:} Performance gains in fine-tuning tasks when using models pretrained on ImageNet, compared to those trained from scratch.

\textbf{Validity of Claim 2 from Evidence:}
\begin{enumerate}
    \item \textbf{Content Validity \warnemoji}
    \begin{itemize}
        \item \textit{Strength:} The wide coverage of natural image phenomena—including fine-grained details and numerous object classes—supports the learning of varied and versatile visual features.
        \item \textit{Weakness:} It may not comprehensively represent features present in non-natural or synthetic environments, nor fully capture abstract contextual cues.
        \item \textit{Suggestions:} Consider integrating supplementary datasets that include synthetic, non-natural, or contextually complex images to achieve a more comprehensive assessment.
    \end{itemize}
    \item \textbf{Criterion Validity \okemoji}
    \begin{itemize}
        \item \textit{Strength:} Empirical studies (e.g.,~\cite{Kornblith2018-ho}) show that ImageNet pretraining is strongly predictive of improved fine-tuning and transfer learning outcomes and that performance is concurrent with established classification tasks, addressing both the predictive and concurrent dimensions.
        \item \textit{Weakness:} Although the predictive correlation is robust, direct and extensive concurrent comparisons with alternative feature assessment methods are less common.
        \item \textit{Suggestions:} Enhance validation by conducting side-by-side evaluations comparing learned features across different pretraining methods and downstream tasks.
    \end{itemize}
    \item \textbf{Construct Validity \warnemoji}
    \begin{itemize}
        \item \textit{Strength:} The improvement in fine-tuning performance suggests that the learned features are semantically rich and transferable. This provides evidence of structural validity (as features capture fundamental visual components), convergent validity (via correlation with downstream task performance), and discriminant validity (in differentiating meaningful features from noise).
        \item \textit{Weakness:} It is challenging to definitively establish that these benefits are due to genuine generalization of visual features rather than overfitting to ImageNet-specific patterns, leaving the discriminant aspect less clear.
        \item \textit{Suggestions:} Continually perform in-depth analyses—such as saliency mapping or kernel visualization—to further elucidate the nature of the learned features and clarify the extent of structural, convergent, and discriminant validity~\citep{Simonyan2013-xy}.
    \end{itemize}
    \item \textbf{External Validity \warnemoji}
    \begin{itemize}
        \item \textit{Strength:} The benefits of ImageNet pretraining have been observed across multiple downstream benchmarks, suggesting that the learned features generalize beyond the confines of natural images~\citep{Kornblith2018-ho, he2020momentum}.
        \item \textit{Weakness:} The degree of generalizability across the span of domains (e.g., synthetic or non-natural images) remains to be fully validated.
        \item \textit{Suggestions:} Broaden external validation by pretraining on a more diverse set of data and assessing performance on cross-domain tasks.
    \end{itemize}
    \item \textbf{Consequential Validity \warnemoji}
    \begin{itemize}
        \item \textit{Strength:} The transformative impact of ImageNet pretraining in advancing computer vision is well-documented, highlighting its practical benefits.
        \item \textit{Weakness:} An overreliance on fine-tuning improvements may obscure limitations in the intrinsic quality of the learned features, risking overgeneralization regarding model capability.
        \item \textit{Suggestions:} Clearly communicate that fine-tuning gains indicate enhanced performance in specific settings rather than a comprehensive measure of visual feature quality; encourage complementary evaluations focused specifically on feature robustness.
    \end{itemize}
\end{enumerate}

\bigskip

\textbf{Object of Claim:} Visual understanding.\\
\textbf{Claim 3:} ImageNet provides an indication of a model’s overall visual understanding beyond simple label prediction or isolated feature representation.\\
\textbf{Evidence:} Performance on the standard classification task under controlled evaluation conditions, independent of training context.

\textbf{Validity of Claim 3 from Evidence:}
\begin{enumerate}
    \item \textbf{Content Validity \xemoji}
    \begin{itemize}
        \item \textit{Strength:} The task of image classification is well-defined and widely used as a proxy for certain aspects of visual understanding.
        \item \textit{Weakness:} Relying solely on classification does not capture the full range of visual understanding, which includes spatial reasoning, object detection, contextual awareness, and causal interpretation. Understanding is multitask, including detection, segmentation, etc., which are not sufficiently investigated.
        \item \textit{Suggestions:} Complement the classification task with additional evaluations—such as object detection, visual question answering, or spatial reasoning challenges—to more fully capture the construct.
    \end{itemize}
    \item \textbf{Criterion Validity \xemoji}
    \begin{itemize}
        \item \textit{Strength:} Classification accuracy is a clear and quantifiable metric that enables direct comparison across models, addressing both predictive and concurrent aspects to some degree.
        \item \textit{Weakness:} There is limited evidence that high performance on this narrow task reliably predicts the broader and deeper aspects of overall visual understanding.
        \item \textit{Suggestions:} Compare ImageNet classification results with those from benchmarks explicitly designed to evaluate advanced visual reasoning and interpretative skills.
    \end{itemize}
    \item \textbf{Construct Validity \xemoji}
    \begin{itemize}
        \item \textit{Strength:} Operationalizing visual understanding as performance on image labeling provides a measurable framework that reflects a basic structural organization of visual recognition. However, it offers only limited convergent evidence with tasks requiring integrated reasoning and does not fully differentiate (discriminant validity) between mere pattern recognition and comprehensive understanding.
        \item \textit{Weakness:} This narrow operational approach may oversimplify the construct, favoring models that exploit dataset biases rather than achieving holistic visual comprehension.
        \item \textit{Suggestions:} Introduce complementary evaluation tasks (e.g., visual question answering or spatial reasoning challenges) to capture additional dimensions of visual understanding and enhance assessments of structural, convergent, and discriminant validity.
    \end{itemize}
    \item \textbf{External Validity \xemoji}
    \begin{itemize}
        \item \textit{Strength:} ImageNet’s evaluation framework is reproducible, and similar performance trends have been observed across related image-based tasks.
        \item \textit{Weakness:} Its ability to generalize to tasks requiring integrated reasoning, spatial awareness, and contextual interpretation remains unconfirmed.
        \item \textit{Suggestions:} Validate the broader aspects of visual understanding by employing a wider array of benchmarks that emphasize multidimensional reasoning and contextual evaluation.
    \end{itemize}
    \item \textbf{Consequential Validity \xemoji}
    \begin{itemize}
        \item \textit{Strength:} The benchmark has stimulated important discussions on the limitations of measuring visual intelligence solely via classification, underscoring the need for more comprehensive evaluation methods.
        \item \textit{Weakness:} High classification accuracy might be erroneously interpreted as evidence of complete visual understanding, potentially misleading real-world applications.
        \item \textit{Suggestions:} Provide clear guidelines on the interpretative scope of ImageNet results and promote complementary measures to capture the full spectrum of visual intelligence.
    \end{itemize}
\end{enumerate}

\end{document}